\documentclass[useAMS,usenatbib]{./mn2e}
\usepackage{graphicx}
\usepackage{times}
\usepackage{amsmath}
\usepackage{amsfonts}
\usepackage{amssymb}
\usepackage{multirow}
\usepackage{longtable}
\usepackage{afterpage}
\usepackage{pdflscape}

\RequirePackage{color}

\title[Fundamental properties of stars]{
Fundamental properties of stars from {\it Kepler} and {{\it Gaia}} data:
 parallax offset and revised scaling relations 
}
\author[M. Y\i ld\i z and S. \"Ortel ]{M. Y\i ld\i z$^{}$\thanks{E-mail:
mutlu.yildiz@ege.edu.tr} and S. \"Ortel$^{}$\\  
$ $\\
$^{}$Department of Astronomy and Space Sciences, Science Faculty, Ege University, 35100, Bornova, \.Izmir, Turkey\\
}

\begin{document}
\date{Accepted 2021 April 6. Received 2021 April 6; in original form 2019 November 19}

\pagerange{\pageref{firstpage}--\pageref{lastpage}} \pubyear{2019}
\def\braket#1{\left<#1\right>}
\newcommand{\yildiz}{Y\i ld\i z }
\newcommand{\etal}{et al. }
\newcommand{\wrt}{with respect to }
\newcommand{\logg}{\log(g) }
\newcommand{\numino}{\mbox{\ifmmode{\overline{\nu_{\rm min}}}\else$\overline{\nu_{\rm min}}$\fi}}
\newcommand{\numin}{\mbox{\ifmmode{\nu_{\rm min}}\else$\nu_{\rm min}$\fi}}
\newcommand{\teff}{\mbox{\ifmmode{T_{\rm eff}}\else$T_{\rm eff}$\fi}}
\newcommand{\teffsun}{\mbox{\ifmmode{{\rm T}_{\rm eff{\odot}}}\else${\rm T}_{\rm eff{\odot}}$\fi}}
\newcommand{\numax}{\mbox{$\nu_{\rm max}$}}
\newcommand{\nuH}{\mbox{\ifmmode{\nu_{\rm minH}}\else$\nu_{\rm minH}$\fi}}
\newcommand{\nuL}{\mbox{\ifmmode{\nu_{\rm minL}}\else$\nu_{\rm minL}$\fi}}
\newcommand{\Dnu}{\mbox{$\Delta \nu$}}
\newcommand{\muHz}{\mbox{$\mu$Hz}}
\newcommand{\kepler}{\mbox{{\it Kepler}}}
\newcommand{\corot}{\mbox{{\it CoRoT}}}
\newcommand{\numaxS}{\mbox{$\nu_{\rm max {\odot}}$}}
\newcommand{\MS}{{\rm M}\ifmmode_{\odot}\else$_{\odot}$~\fi}
\newcommand{\RS}{{\rm R}\ifmmode_{\odot}\else$_{\odot}$~\fi}
\newcommand{\LS}{{\rm L}\ifmmode_{\odot}\else$_{\odot}$~\fi}
\newcommand{\MSbit}{{\rm M}\ifmmode_{\odot}\else$_{\odot}$\fi}
\newcommand{\RSbit}{{\rm R}\ifmmode_{\odot}\else$_{\odot}$\fi}
\newcommand{\LSbit}{{\rm L}\ifmmode_{\odot}\else$_{\odot}$\fi}
\maketitle
\label{firstpage}
\begin{abstract}

{
Data from the space missions {\it Gaia}, {\it Kepler}, {\it CoRoT} and {\it TESS}, make it possible to compare parallax and asteroseismic distances. 
From the ratio of two densities $\rho _{\rm sca}/\rho_{\pi}$, we obtain an empirical relation $f_{\Delta \nu}$ between 
the asteroseismic large frequency separation and mean density, which is important for more accurate stellar mass and radius. This expression for  
main-sequence (MS) and subgiant stars with $K$-band magnitude is very close to the one obtained from interior MS models
by  Y{\i}ld{\i}z,  \c{C}elik \& Kayhan. We also discuss the effects of effective temperature and parallax offset as the source of
the difference between asteroseismic and non-asteroseismic stellar parameters.
We have obtained our best results for about 3500 red giants (RGs) by using 2MASS data and model values for $f_{\Delta \nu}$ from Sharma et al.
Another unknown scaling parameter $f_{\nu_{\rm max}}$ comes from  the relationship between the frequency of maximum amplitude and gravity.
Using different combinations of $f_{\nu_{\rm max}}$ and the parallax offset, we find that the parallax offset is generally a function of distance.  
The situation where this slope disappears is accepted as the most reasonable solution.
By a very careful comparison of asteroseismic and non-asteroseismic parameters, we obtain very precise values for the parallax offset 
and $f_{\nu_{\rm max}}$ for RGs of $-0.0463\pm0.0007$ mas and $1.003\pm0.001$, respectively.
Our results for mass and radius are in perfect agreement with those of APOKASC-2:  the mass and radius of $\sim$3500 RGs are in the 
range of about 0.8-1.8 M$_{\odot}$ (96 per cent) and 3.8-38 R$_{\odot}$, respectively. 
}
\end{abstract}

\begin{keywords}
stars: distance -- stars: evolution -- stars: fundamental parameters -- stars: interiors -- stars: late-type -- stars: oscillations.
\end{keywords}

\section{Introduction}
The determination of stellar properties is very important for our understanding of astronomical structures from planets to galaxies.
Thanks to the {\it Kepler} (Borucki et al. 2010), {\it CoRoT} (Baglin et al. 2006)  and {\it TESS} (Sullivan et al.  2015) space telescopes, 
asteroseismic scaling relations can be used to estimate fundamental stellar parameters.  One of the essential parameters that we require to determine
fundamental stellar properties or to check the relations between 
these properties is the distance of stars from us. 
We have already 
compared asteroseismic distances $d_{\rm sca}$ derived from the so-called asteroseismic scaling relations with  the {\it Gaia} distances
$d_{\rm DR1}$ from Data Release 1 (DR1; Brown et al. 2016; Y\i ld\i z et al. 2017). 
There was good agreement between them for the 64 stars, 
with a difference of less than 10 per cent. 
In the present study, we compare $d_{\rm sca}$ and the recent {\it Gaia} distances $d_{\pi}$ from DR2 (Brown  et  al.  2018) 
and we try to improve asteroseismic relations,
also using the data of solar-like oscillating stars in eclipsing binaries (EBs; Gaulme et al. 2016) and first-ascent red giants (RGs) in APOKASC-2  (Pinsonneault, Elsworth \& Tayar 2018). The total number of stars analysed is about 3600. The red clump (RC) stars in APOKASC-2 are not included in our analysis.

The {\it Gaia} parallax zero-point offset has been widely discussed in the literature (e.g. Lindegren et al. 2018; Leung \& Bovy 2019; Zinn et al. 2019a).
We also discuss the parallax zero-point offset with and without improved scaling relations.

For the computation of the asteroseismic distance of a star, the most important outcome of asteroseismology is the scaling relation for radius, $R_{\rm sca}$. 
This relates the radius to the so-called large frequency separation $\braket{\Dnu}$ ({Tassoul 1980, Christensen-Dalsgaard 1993}),
the frequency of maximum amplitude \numax ({Brown et al. 1991; Kjeldsen \&
Bedding 1995}) 
and the effective temperature \teff: 
\begin{equation}
\frac{R_{\rm sca}}{\rm R_{\odot}}=\frac{\numax/\nu_{\rm max\odot}}{(\braket{\Delta \nu}/\braket{\Delta \nu_\odot})^2}\left( \frac{T_{\rm eff}}{\rm T_{\rm eff\odot}}
\right)^{1/2}.
\end{equation}
Here, ${\rm R_{\odot}}$, ${\rm T_{\rm eff\odot}}$, { $\nu_{\rm max\odot}$ and $\braket{\Delta \nu_\odot}$ are the solar values of radius, effective temperature, $\numax$,} and $\braket{\Delta \nu}$, respectively.
Then, using $R_{\rm sca}$ and spectroscopic \teff, we can obtain the luminosity $L_{\rm sca}$ and the bolometric magnitude.
By computing a bolometric correction, $BC$, from the tables, we can utilize the distance modulus and  hence infer the star's distance, $d_{\rm sca}$.
If we have highly precise parallaxes,
the comparison of $d_{\rm sca}$ and $d_{\pi}$ will 
clarify how accurate the asteroseismic relation for $R$ is, and how accurate the spectroscopic measurement of \teff~ is.

For the computation of stellar mass from $d_{\pi}$, the surface gravity of stars, $\log g$, should also be very precisely determined.  
The exact form of $M_{\pi}$ is given as 
\begin{equation}
\frac{M_\pi}{\rm M_{\odot}}= \frac{g/g_{\odot}}{(\teff/\teff_{\odot})^{4}}10^{-[V-A_V+5(1-\log d_\pi)+BC-{\rm M}_{\rm bol\odot}]/2.5},
\end{equation}
in solar units, where $V$, $A_V$, ${\rm M_{\odot}}$ and ${\rm M}_{\rm bol\odot}$ are apparent magnitude, 
interstellar extinction, the solar mass  and the solar bolometric magnitude, respectively, and
$d_\pi$ is in units of pc.
Equation (2) is a combination of five different equations. Four equations are the expressions for the absolute magnitude, the bolometric magnitude, the gravity and the luminosity. 
The fifth equation is the relation between the bolometric magnitude and the luminosity.
The standard scaling relation for mass $M_{\rm sca}$, however, is given as
\begin{equation}
\frac{M_{\rm sca}}{\rm M_{\odot}}=\frac{(\numax/\nu_{\rm max\odot})^3}{(\braket{\Delta \nu}/\braket{\Delta \nu_\odot})^4}\left( \frac{T_{\rm eff}}{\rm T_{\rm eff\odot}}
\right)^{3/2}.
\end{equation}

It is well known that sound speed is a function of compressibility (the first adiabatic exponent, $\Gamma_{\rm \negthinspace 1}$), in addition to temperature.
In deriving scaling relations between asteroseismic and non-asteroseismic quantities, compressibility at the stellar surfaces, $\Gamma_{\rm \negthinspace 1s}$, is taken as a constant.
However, Y\i ld\i z,  \c{C}elik Orhan \& Kayhan (2016) have shown that $\Gamma_{\rm \negthinspace 1s}$ is not constant at the stellar surfaces and they modified the expressions for 
\numax~ and $\Delta \nu$ as functions of $\Gamma_{\rm \negthinspace 1s}$. They also found how compressibility depends on $T_{\rm eff}$ for main-sequence (MS) stars. 
A similar modification of the scaling relation may be required as a function of mean molecular weight as well (Viani et al. 2017).


De Ridder et al. (2016) generally find very good agreement (and a significant parallax offset) between $d_{\rm sca}$ and $d_{\rm DR1}$ for 22 dwarfs and subgiant solar-like oscillators. 
They also confirm that $d_{\rm DR1}$ is more uncertain than $d_{\rm sca}$ for  a sample of 938 RG solar-like oscillators ($d_{\rm sca}$ $>$ 300 pc), more distant than the 
pulsating dwarfs ($d_{\rm sca}$ $<$ 250 pc). 
Davies, Lund \& Miglio (2017) compare the {\it Gaia} DR1 distances with $d_{\rm sca}$ of RC stars and the distances obtained using luminosity
determined by EBs. They find that $d_{\rm sca}$ is in better agreement with the RC distance estimators than the DR1 distances.  
They propose a correction to $d_{\rm DR1}$ as a function of distances $>$ 500 pc for the {\it Kepler} field of view.
Huber et al. (2017) compare parallaxes and radii from asteroseismology and {\it Gaia} DR1 
for 2200 {\it Kepler} stars spanning from the MS to the red giant branch. 
They find a distance offset no larger than 5 per cent for MS stars and low-luminosity RGs.


Stassun \& Torres (2018) used eclipsing binary distances to compare the distances from the {\it Gaia} DR2 parallaxes. They found a mean offset
of $-82 \pm 33$ $\mu$as, in the sense that the {\it Gaia} parallaxes are too small. A similar offset ($-52.8 \pm 2.4$ $\mu$as) is identified by 
Zinn  et al. (2019a) for the RGs from the APOKASC-2 catalogue. 
The {\it Gaia} team finds that the offset is dependent  on color and magnitude. While the global mean offset is statistically determined as $-0.029$ mas in Lindegren et al. (2018),
the zero-point offset  from  0.09 to $-0.118$ mas is reported by Arenou et al. (2018) for various sources.
Zinn et al. (2019b), Khan et al. (2019), Hall et al. (2019) and Sahlholdt et al. (2018) try to revise conventional asteroseismic 
scaling relations in order to obtain consistency between $d_{\rm sca}$ and $d_{\pi}$.
We show how the parallax zero-point offset and improvements in scaling relations are inter-related (see Section 5).



This paper is organized as follows.
In Section 2, we present DR2 distances and other observational data for about 3600 stars, and we compare DR1 and DR2 distances of 74 stars 
 (Y17 sample) studied in Y\i ld\i z et al. (2017).
In Section 3, we briefly explain the method for the computation of asteroseismic distance.
Section 4 is devoted to the results and their comparison.
The parallax offset issue is considered in Section 5.
Finally, in Section 6, we draw our conclusions.

\section{{ \it GAIA} DISTANCES: COMPARISON OF DR1 AND DR2 DATA}
We have studied 89 solar-like oscillating stars in Y\i ld\i z et al. (2016, 2019).
In Y\i ld\i z et al. (2016), it is shown that scaling relation between the mean large separation and mean density depends on 
$\Gamma_{\rm \negthinspace 1s}$ and the reference frequencies (\numin$_0$, \numin$_1$ and \numin$_2$) derived from glitches due to
the He {\scriptsize II} ionization zone have very
strong diagnostic potential for the determination of \teff. 
In Y\i ld\i z et al. (2019), the same stars are analysed and their mass, radius and age are computed from different scaling relations including 
relations based on \numin$_0$, \numin$_1$ and \numin$_2$.
Parallaxes of the Y17 sample are available in DR1. Their $d_{\rm sca}$ and $d_{\rm DR1}$ are already compared in Y\i ld\i z et al. (2017).
Their distances range 21-433 pc.
\begin{figure}
\includegraphics[width=101mm,angle=0]{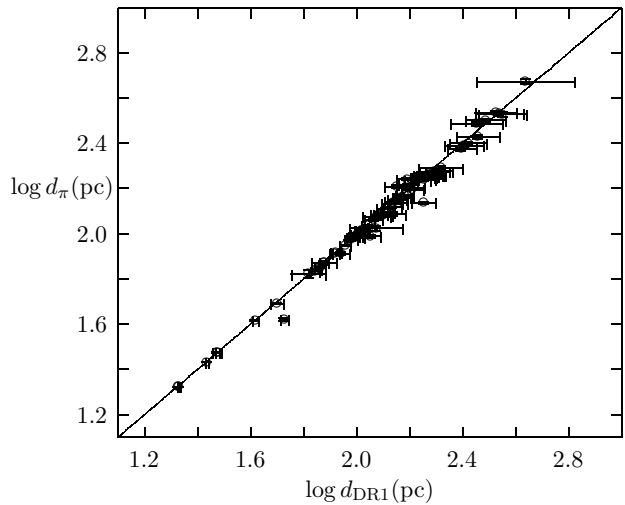}
\caption{The figure plots $\log d_{\pi}$ of the Y17 stars \wrt $\log d_{\rm DR1}$ in units of pc. 
}
\end{figure}

In Fig. 1, $\log d_{\pi}$ of the Y17  stars is plotted \wrt $\log d_{\rm DR1}$, with their uncertainties.
The difference between $d_{\pi}$ and $d_{\rm DR1}$ 
is less than 4 per cent for 54 stars and greater than 10 per cent for seven stars.
The mean difference is about  2.4 per cent. The greatest differences occur for KIC 8379927/HIP 97321 and KIC 1435467. 
The differences between their $d_{\rm DR1}$ and $d_{\pi}$ are 27 and 30 per cent, respectively.
For five stars, namely,  
KIC 4914923/HIP 94734, KIC 6933899, KIC 12317678/HIP 97316, KIC 12508433 and  HD 181907/HIP 95222,
the difference between the distances is about 9-13 per cent.

We notice that $d_{\pi}$ is much more precise than $d_{\rm DR1}$. 
In Fig. 2, uncertainty in $d_{\pi}$ ($\Delta d_{\pi}$)  is plotted \wrt uncertainty in $d_{\rm DR1}$ ($\Delta d_{\rm DR1}$)
in the log-log plot.  The solid line is the fitted line.  The implication is that $\Delta d_{\pi}$ is about one-tenth of $\Delta d_{\rm DR1}$. 
However, it is reported in Brown et al. (2018) that the astrometric uncertainties may be underestimated (see also Arenou et al. 2018). 
\begin{figure}
\includegraphics[width=101mm,angle=0]{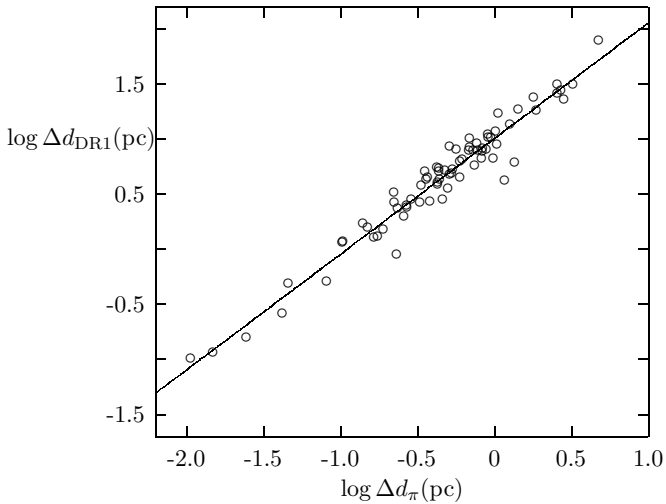}
\caption{The figure plots $\log \Delta d_{\rm DR1}$ of the Y17 stars \wrt $\log \Delta d_{\pi}$ in units of pc. 
The solid line is the fitted line of $\log \Delta d_{\rm DR1}=1.052\log \Delta d_{\pi}+1.007$. This implies that $d_{\pi}$ is about 10 times 
more precise than $d_{\rm DR1}$.  
}
\end{figure}
 
In addition to the Y17 stars, the parallaxes of five more solar-like oscillating stars studied in Y\i ld\i z et al. (2016) are available in DR2.  These stars are KIC 8219268, KIC 11026764,   
HD 43587/HIP 29860, HD 146233/HIP 79672 and HD 203608/HIP 105858. 
The latter three stars are relatively very close to the Sun, the closest stars among the 79 stars (Y16 DR2 sample). Their distances are 19.30, 14.13 and 9.27 pc, respectively.
Their solar-like oscillations were detected from {\it CoRoT} light curves.
KIC 8219268 and  KIC 11026764 were among the {\it Kepler} targets. $d_{\pi}$ of KIC 11026764 is 156.89 pc, while KIC 8219268 is 1345 pc away and 
the farthest of the Y16 DR2 sample. It is a red giant and its radius is about 6.7 \RSbit. 
{The references for  \teff, [Fe/H] and asteroseismic data for Y16 DR2 are given in Y\i ld\i z et al. (2019).}

{The motivation behind this study is that the parallaxes given in DR2 are much more precise than in DR1. 
This may allow us to reach more conclusive results for the Y16 DR2 sample. 
The analysis of this sample} leads us to extend this study to include RGs. In addition to the solar-like oscillating RGs in eclipsing binaries,
we also consider 909 RGs from {APOKASC-2} for which the {\it Gaia} DR2 parallax and $V$ are available {(VRG sample)}. 
{The spectroscopic parameters in APOKASC-2 { (Nidever et al. 2015; Holtzman et al. 2015, 2018; Garc{\'\i}a P{\'e}rez et al. 2016; Wilson et al. 2019)} are taken from the 14th data release 
of the Sloan Digital Sky Survey (Gunn et al. 2006; Abolfathi et al. 2018).}
In our analysis, we use effective temperatures $T_{\rm A}$ given in APOKASC-2.
The distances of the stars are in the range  200-11 250 pc. While most of {the VRG} stars have a $d_{\pi}$ less than 4000 pc, there are only 10 giants with $d_{\pi}>$ 6000 pc.
$V$ {and $K$ of these RGs are} taken from the SIMBAD data base, compiled from different sources. For more uniform data, we also use $G$ of the RGs from the {\it Gaia} 
data base {(GRG sample)}. The number of stars for which the interstellar extinction $A_g$ is available, which are also present in APOKASC-2, is 1323.  

{In addition to $V$ and $G$ magnitudes, we also employ $J$, $H$ and $K$ magnitudes (JHKRG sample) from the 2MASS ({Skrutskie et al. 2006}).
This sample consists of 3524 RGs and has the largest number of stars among the samples. 
There are very important advantages to using this data set, perhaps the most important being that interstellar extinction for $J$, $H$ and $K$ is significantly
lower than extinction for $V$ and $G$.  
}


Using the distance of stars and $V$, we can compute absolute magnitude and then their luminosity $L_{\pi}$. For the computation of luminosity,   $BC$ is 
derived by interpolating between the colour-$T_{\rm eff}$ tables of Lejeune, Cuisinier \& Buser (1998) or the MESA isochrones and stellar tracks ({MIST; Paxton et al. 2011, 2013;} Choi et al. 2016). 
As we have the $T_{\rm eff}$ of stars from their spectroscopic observations, 
the radius of stars $R_{\pi}$ can be obtained from $ R_{\pi}=[L_{\pi}/(4\pi \sigma T_{\rm eff}^4)]^{0.5}$,
where $\sigma$ is the Stefan-Boltzmann constant. 
Using either spectroscopic $\log g_{\rm S} $ or 
asteroseismic $\log g_{\rm sca}$ gravity, stellar mass can also be {derived}. For a reliable stellar mass, in addition to distance with a small uncertainty, a very precise $\log g$ value is required.

{The four samples of stars we analyse (i.e. the Y16 DR2, VRG, GRG and JHKRG samples) are plotted in Fig. 3.} There are few RGs in Y16 DR2, half of the rest are MS stars and 
the other half are in subgiant (SG) phase.
{For  Y16 DR2, both $V$- and $K$-band magnitudes are used.}
\begin{figure}
\includegraphics[width=101mm,angle=0]{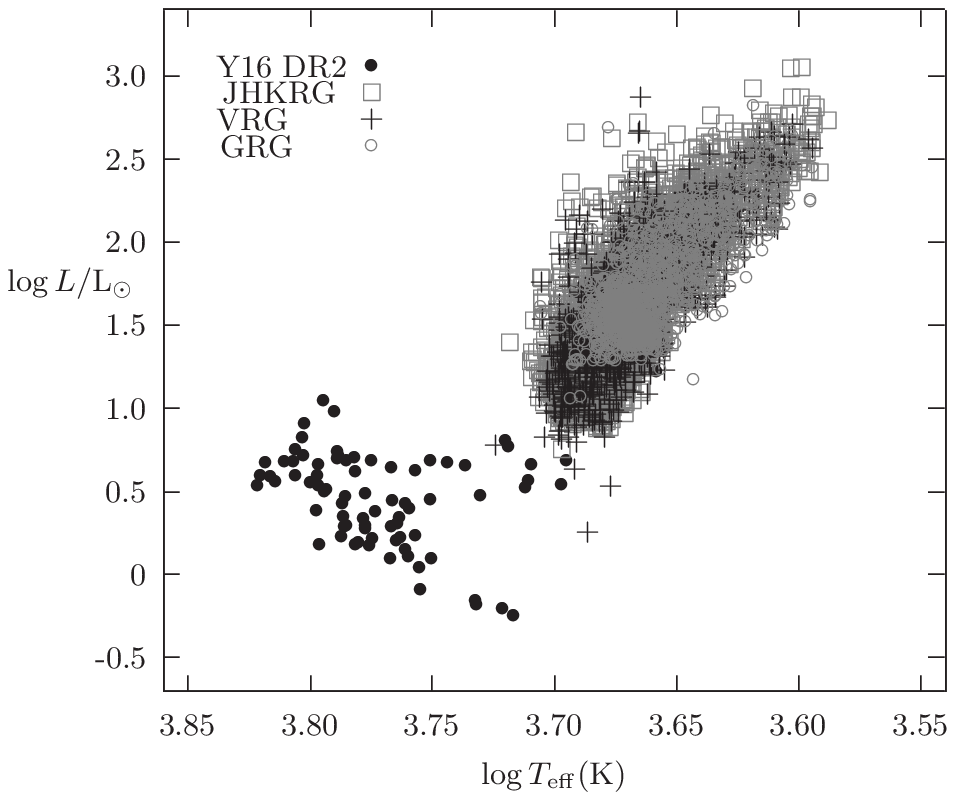}
\caption{The figure plots $\log L/{\rm L}_\odot$ of four samples \wrt $\log \teff$.
The filled circles, squares, crosses and open circles denote Y16 DR2, JHKRG, VRG and GRG, respectively.
}
\end{figure}

{Interstellar extinction for the $V$-band, namely $A_V$, is computed from $E(B-V)$ from Schlegel, Finkbeiner \& Davis (1998), as recalibrated by Schlafly \& Finkbeiner (2011):  
$A_V=3.1 E(B-V)$.
$A_G$ for the $G$-band magnitude is already given in the {\it Gaia} data base for the GRG sample stars. 
The extinction coefficient for the $K$ magnitude, $A_K$, is obtained as $A_K= 0.287  E(B-V)$ by Fulton \& Petigura (2018).
We take the extinction coefficients for $H$ and $J$ magnitudes as $A_H= 1.74 A_K$ (Nishiyama et al. 2008) and $A_J= 2.7 A_K$ (Martin \& Whittet 1990), respectively. 
}

\section{ASTEROSEISMIC DISTANCES, MASSES AND RADII}

We compute $d_{\rm sca}$ and the distance uncertainty $\Delta d_{\rm sca}$ of the solar-like oscillating stars as in Y\i ld\i z et al. (2017).
In short, in order to compute the distance modulus from the absolute magnitude, we must first obtain luminosity. 
The luminosity of these stars is computed using $T_{\rm eff}$ and $R_{\rm sca}$.
In addition to these quantities, visual magnitude $V$ (or $G$), metallicity ([Fe/H])  and bolometric correction $BC$ are required to
compute $d_{\rm sca}$ of a star.
We use both the conventional {(equations 1 and 3) and  modified  (see below)}
asteroseismic scaling relations to obtain the mass and radius of the stars.

The scaling relations for $M$ and $R$ are based on two expressions for $\Delta \nu$ and \numax.
{The expression for $\Delta \nu$ is}
\begin{equation}
\frac{\braket{\Dnu}}{\braket{\Dnu_{\odot}}}=f_{\Delta \nu}\frac{\braket{\rho}^{1/2}}{\braket{\rho_{\odot}}^{1/2}},
\end{equation}
{where $f_{\Delta \nu}$ (White et al. 2011; Sharma et al. 2016; Y{\i}ld{\i}z et al. 2016) encapsulates errors in the assumption of homology and also may capture issues with the surface correction.
In equation (4), $\braket{\rho}$ and $\braket{\rho_{\odot}}$ are the mean stellar and solar density, respectively.}
Similarly,
\begin{equation}
\frac{\numax}{\numax_{\odot}}= f_{\nu_{\rm max}}\frac{g/g_{\odot}}{(\teff/\teff_{\odot})^{1/2}},
\end{equation}
{where $g_{\odot}$ in equation (5) is the solar surface gravity.} 
{The parameter $f_{\nu_{\rm max}}$, as $f_{\Delta \nu}$, also encapsulates errors in the assumption of homology and also may capture issues with the surface correction. }

$f_{\nu_{\rm max}}$ is related to $f_{\nu}$ introduced by Y\i ld\i z et al. (2016) as
\begin{equation}
f_{\nu_{\rm max}}=f_{\nu} \left(\frac{\mu_{\rm s} \Gamma_{\rm \negthinspace 1s}}{\mu_{\rm s \odot} \Gamma_{\rm \negthinspace 1s{\odot}}}\right)^{1/2}
\end{equation}
{where $\Gamma_{\rm \negthinspace 1s}$ and $\mu_{\rm s}$ are the compressibility and  the mean molecular weight at the stellar surface, respectively}, and 
$\Gamma_{\rm \negthinspace 1s{\odot}}$ and  $\mu_{\rm s \odot}$ are compressibility and the mean molecular 
weight at the solar surface, respectively.
The compressibility is mainly a function of \teff~ and is obtained as 
\begin{equation}
\frac{1}{\Gamma_{\rm \negthinspace 1s}}= 1.6 \left(\frac{\teff}{\teff_\odot}-0.96\right)^2+0.607
\nonumber
\end{equation}
{for the interior models of MS stars in Y{\i}ld{\i}z et al. (2016).  
The solar values of $\Gamma_{\rm \negthinspace 1s{\odot}}$, $\braket{\Dnu_{\odot}}$ are taken as $1.639$ and $136.0$ in Y{\i}ld{\i}z et al. (2016). 
This value of $\braket{\Dnu_{\odot}}$ was preferred because it makes $f_{\Delta \nu}=1$ for the Sun, as it should be.
In this study, we take solar values of \Dnu~ and $\nu_{\rm max}$ as  135.1 and  3090 \muHz ~(Sharma et al. 2016), respectively.}
The solar density and gravity are computed from the solar data (Sackmann, Boothroyd \& Kraemer 1993) as 1.4086 g cm$^{-3}$ and 27 403 g cm$^{-2}$, respectively.

The conventional scaling relations assume unity for $f_{\nu_{\rm max}}$ and $f_{\Delta \nu}$. For the solar-like oscillating MS stars, this can be 
a good approach. However, the structure of evolved stars is significantly different from the solar structure. 
{For the non-standard scaling relation for radius $R'_{\rm sca}$, 
$R_{\rm sca}$ must be multiplied by $f_{\Delta \nu}^2/f_{\nu_{\rm max}}$:
\begin{equation}
\frac{R'_{\rm sca}}{\rm R_{\odot}}=\frac{\numax/\nu_{\rm max\odot}}{(\braket{\Delta \nu}/\braket{\Delta \nu_\odot})^2}\left( \frac{T_{\rm eff}}{\rm T_{\rm eff\odot}}
\right)^{1/2}
\frac{f_{\Delta \nu}^2}{f_{\nu_{\rm max}}}.
\end{equation}
For the non-standard scaling relation for mass $M'_{\rm sca}$, the factor is $f_{\Delta \nu}^4/f_{\nu_{\rm max}}^3$:
\begin{equation}
\frac{M'_{\rm sca}}{\rm M_{\odot}}=\frac{(\numax/\nu_{\rm max\odot})^3}{(\braket{\Delta \nu}/\braket{\Delta \nu_\odot})^4}\left( \frac{T_{\rm eff}}{\rm T_{\rm eff\odot}}
\right)^{3/2}
\frac{f_{\Delta \nu}^4}{f_{\nu_{\rm max}}^3}.
\end{equation}
}

Most of the stars in the {APOKASC-2} catalogue are evolved RG stars. Their high brightness allows detection of their solar-like oscillations by the {\it Kepler} and {\it CoRoT}
telescopes, despite their {large distances}. Therefore, any improvement in scaling relations for these stars will be a substantial 
advancement not only in stellar astrophysics but also in our understanding of chemical evolution and dynamics of {the Galactic disc}.

\section{Results and Discussions}

\subsection{Comparison of distances of the {Y16 DR2 } stars}
In Fig. 4, we plot {$\log d_{\rm sca}$ \wrt $\log d_{\pi}$ for  the Y16 DR2 {sample with (circles) and without  (filled circles) interstellar extinction.  
The parallax of the five stars is not found in DR1, but is in DR2.}
{The observational \teff~ and metallicities of these {stars are compiled} from the literature (see table B1 in Y\i ld\i z et al. 2019 for the references)}.
There is, in general, very good agreement between $d_{\rm sca}$ and $d_{\pi}$ of the Y16 DR2 sample without extinction. 
The most striking feature of Fig. {4} is to see how successful both asteroseismic and astrometric methods are in determining the
distance of a star at $1345$ pc (KIC 8219268). 
{However, there is a significant difference between $\log d_{\rm sca}$} and $\log d_{\pi}$ if the extinction is included. This implies that
the extinction is very large, in particular for stars with $d_{\pi}<$ 200 pc.}


{If we use the {\it K}-band magnitude in place of the {\it V}-band magnitude, the results we obtain are very similar to those given in Fig. 4 without extinction. The mean difference between $d_{\pi}$ and $d_{\rm sca}$
is about 2.8 per cent for the Y16 DR2 sample. In Y\i ld\i z et al. (2017), the difference between $d_{\rm DR1}$ and $d_{\rm sca}$ was about  
5 per cent. 
In Fig. 5, we plot the histogram of the normalized difference between {\it Gaia} and asteroseismic distances for  {the Y16 DR2 stars with {\it K}- and {\it V}-band magnitudes}.
For the {\it K}-band magnitude, the distribution is very similar to a Gaussian distribution. The results for the {\it V}-band magnitude are complicated and do not show a 
regular Gaussian curve.
}
\begin{figure}
\includegraphics[width=101mm,angle=0]{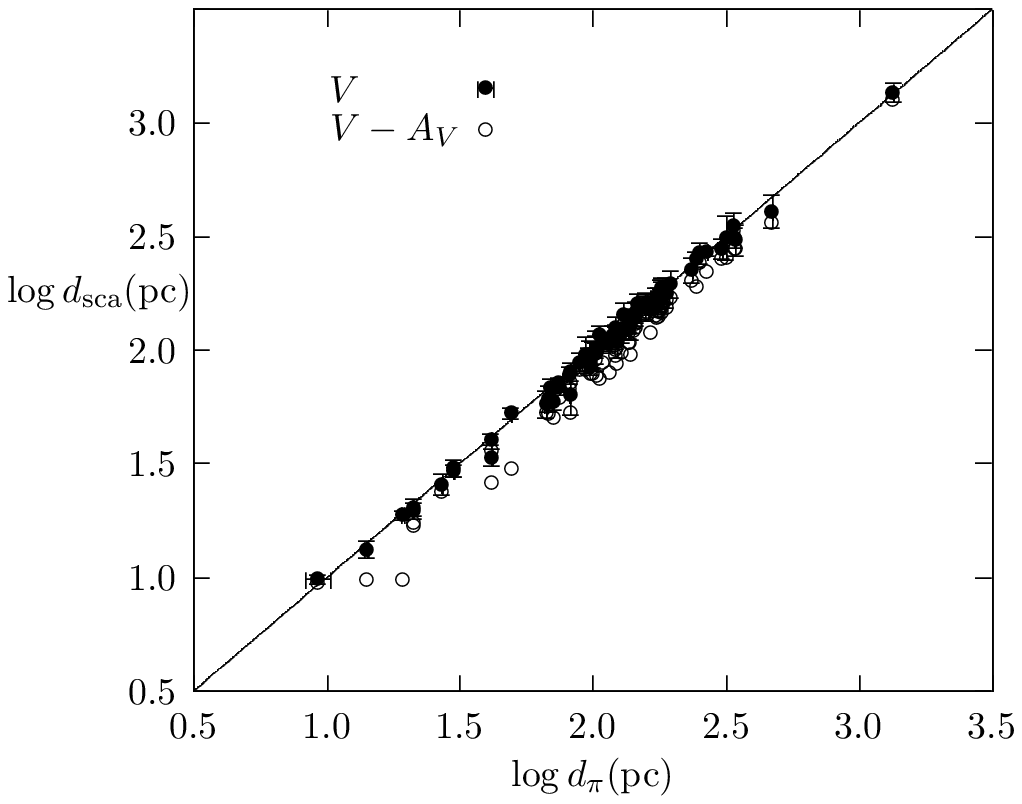}
\caption{The figure plots $\log d_{\rm sca}$ of the Y16 DR2 sample with (circles) and without  (filled circles) interstellar extinction, 
\wrt $\log d_{\pi}$ in units of pc.
}
\end{figure}

\begin{figure}
\includegraphics[width=101mm,angle=0]{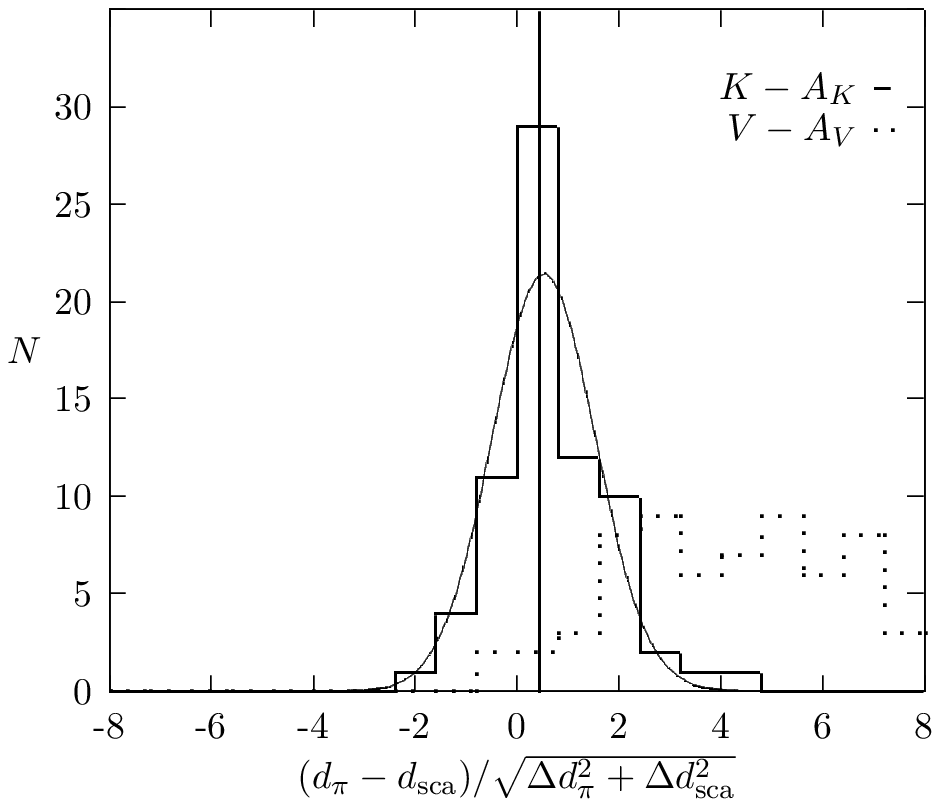}
\caption{
{Histogram of the normalized difference between {\it Gaia} and asteroseismic distances for the Y16 DR2 sample. 
The dotted and solid thick lines are for  $K$- and $V$-band magnitudes, respectively.} 
The vertical line shows the median and the thin line shows the fitted Gaussian curves for the $K$-band magnitude. 
}
\end{figure}
{For 51 stars of the {Y16 DR2 sample}, the difference between  $d_{\pi}$ and $d_{\rm sca}$  {($\delta d=d_{\pi} - d_{\rm sca}$) is less than 6 per cent.}
We notice that distance of most of the stars with $\delta d/d~<$ 0.06 is less than 200 pc.
The largest fractional differences ($\delta d/d$) between  $d_{\pi}$ and $d_{\rm sca}$ are 0.28 and  0.22 for KIC 9025370 and KIC 8379927, respectively.
The difference between the DR1 and DR2 distances of KIC 9025370 is about 6 per cent.
KIC 9025370 has the most uncertain \numax~ ($\Delta \numax=215$ \muHz) among the Y16 DR2 sample. This might be the reason for the difference between the distances.
}

%
%

If we use the modified scaling relation for the radius of stars with $\Gamma_{\rm 1s}$ (equation 10 in Y\i ld\i z et al. 2016 with $f_{\nu_{\rm max}}=1$), then the {mean} difference between 
$d_{\rm sca}$ and $d_{\pi}$ is about 1.8 per cent. 
Although there is a slight difference between 
distances computed from conventional and modified scaling relations, the situation changes if we compare $R_{\rm sca}$ and $R_{\pi}$ (see below).

The $\log g_{\rm S} $ of stars is perhaps the most uncertain parameter derived from spectroscopic observations. We prepare another sample of stars
for which the difference between $\log g_{\rm S} $ and $\log g_{\rm sca} $ is less than 5 per cent. 32 stars are included in this sample. 
For this sample, the {mean} difference between
$d_{\rm sca}$ and $d_{\pi}$ is about 0.4 per cent, much smaller than that of the {Y16 DR2 sample}.
If $|\log g_{\rm S}  - \log g_{\rm sca}   |<0.07$, then the difference in the distance of 42 stars is reduced to 0.1 per cent. Except for seven stars, the difference in the distance for 
all of the stars is less than 6 per cent. {The largest difference occurs for KIC 9025370: $d_{\pi}$=82.0 and $d_{\rm sca}$=$59.0$ pc.}

\subsection{Comparison of radii of the {Y16 DR2 sample} stars}
\begin{figure}
\includegraphics[width=101mm,angle=0]{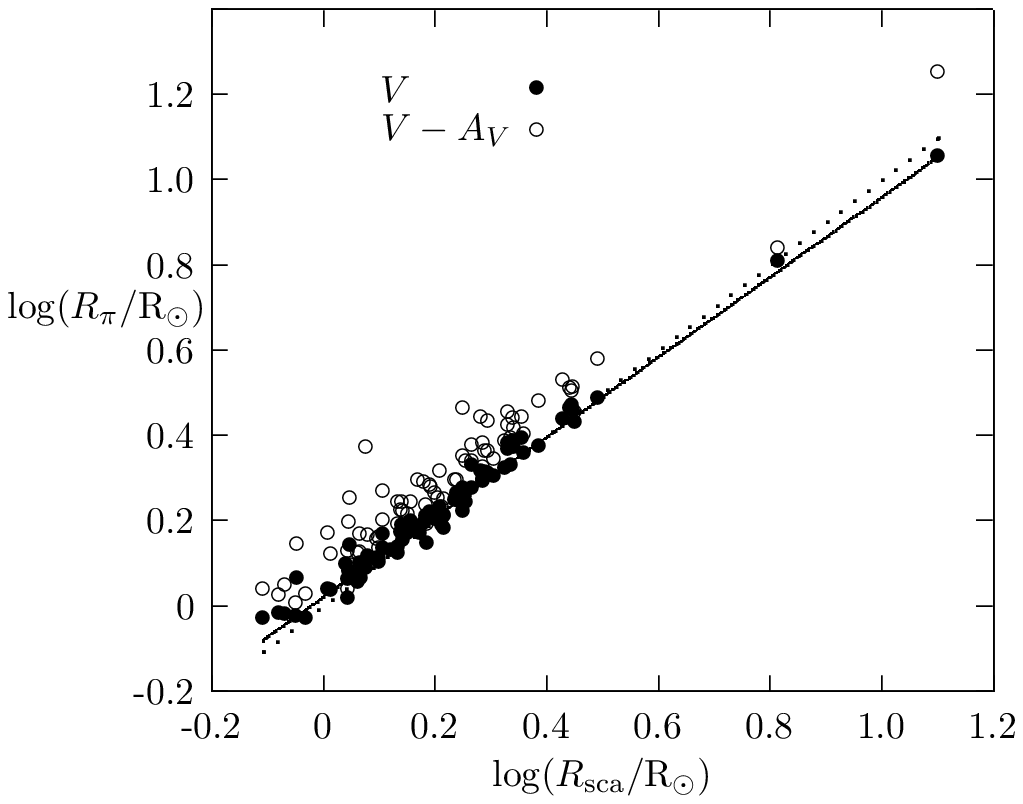}
\caption{The figure plots $\log R_{\pi}$ of {the Y16 DR2 sample \wrt $\log R_{\rm sca}$ in solar units.
{The solid line shows the fitting line  
$\log R_{\pi}=(0.938\pm 0.015)\log R_{\rm sca}+(0.031\pm 0.004)$,}
and the dotted line shows $\log R_{\pi}=\log R_{\rm sca}$.}
}
\end{figure}
{The logarithm of $R_{\pi}$ of the Y16 DR2 sample with and without extinction
is plotted with respect to $\log R_{\rm sca}$ in Fig. 6.
$R_{\pi}$ without extinction is in much better agreement with $R_{\rm sca}$ than $R_{\pi}$ with extinction.
The radii of solar-like oscillating stars are in the range $0.93-11.4$ \RSbit. 
The largest star is HD 181907. It is a red giant and relatively close to us. Its distance is $105.6$ pc and its solar-like 
oscillations were detected by the {\it CoRoT} telescope.
For small radii, $R_{\pi}$ and $R_{\rm sca}$ are in good agreement. For the large values of radii, however, the data 
deviate from the $\log R_{\pi}=\log R_{\rm sca}$ line. {The fitting line  is found to be
$\log R_{\pi}=(0.938\pm 0.015)\log R_{\rm sca}+(0.031\pm 0.004)$; 
hereafter, radii and masses in similar relationships are written in solar units.
This expression is equivalent to $R_{\pi}=1.074R_{\rm sca}^{0.938}$.}
There are clear reasons for having doubts about whether this deviation is real, as there are only two stars with a radius larger than 3.5 \RSbit.
Also, the $\log R_{\pi}=\log R_{\rm sca}$ line and the fitted line apparently deviate after 3.5 \RSbit.
Therefore, we obtain another line by {excluding the two largest stars}. The new line is found to be 
$(0.919\pm 0.024)\log R_{\rm sca}+(0.024\pm 0.006)$.
The two fitted lines are very similar to each other. However, to be sure, the data of the {\it Kepler} RG stars should be analysed (see below).
}  


{If  we use the  $K$-band magnitude, we find that 
$\log R_{\pi}=(0.942\pm 0.018)\log R_{\rm sca}+(0.032\pm 0.005)$. This result is in very good agreement with the fitting line for $V$-band magnitude without extinction.
This result makes us think that either $A_V$ is  excessive for these relatively close stars or the extinction effect is already taken into account in the $V$ data
(see below).
}

In computation of $R_{\rm sca}$ and $R_{\pi}$, we use the same $T_{\rm eff}$, {so $\delta d/d$
is equal to}
$\delta R/R =(R_{\pi}-R_{\rm sca})/R_{\rm sca}$.  The explanation is as following. 
For a given $V$ of a star, we essentially know the ratio of $L_\pi$ to $d^2$. For precise determination of $d$, we must find luminosity. 
For computation of $L_{\rm sca}$, we have spectroscopically or photometrically derived $T_{\rm eff}$ and $R_{\rm sca}$ obtained from 
asteroseismic scaling relations. Because we use the same $T_{\rm eff}$ in the computation of $R_{\rm sca}$ and $R_{\pi}$, the difference between $L_{\rm sca}$ and $L_{\pi}$
is due to the difference between the radii $R_{\rm sca}$ and $R_{\pi}$. In that case, $L/d^2\propto R^2/d^2$. As this ratio is constant for a given $V$,
then $\delta d/d$ is equal to $\delta R/R$.

There is a substantial difference between $R_{\pi}$ and $R_{\rm sca}$. We can correct the asteroseismic radius, $R_{\rm cor}$, according to the fitted line from Fig. 6: 
$R_{\rm cor}=1.074 R_{\rm sca}^{0.938}$, where radii are in solar units.
If we use $R_{\rm cor}$ in place of $R_{\rm sca}$ and recalculate $d_{\rm sca}$, we see that $d_{\rm sca}$ and $d_{\pi}$ are in better agreement. 

Although the correction in $R_{\rm sca}$ improves the agreement between $d_{\rm sca}$ and $d_{\pi}$, $R_{\rm sca}$ deviates from $R_{\pi}$ for only a few stars with large radii.
This deviation can be tested using the RG stars in the {APOKASC-2} catalogue and solar-like oscillating RGs in eclipsing binaries.

\subsection{Solar-like oscillating RGs in EBs and implications for the scaling relations}
For the 10 EBs, the mass $M_{\rm rv}$ and radius $R_{\rm rv}$ of the component stars 
are obtained by analysing their light curves and radial velocities (Gaulme et al. 2016). 
The RG components in these binaries are solar-like oscillating stars.
Their \numax~ and $\braket{\Dnu}$ are derived from the {\it Kepler} light curves.  
These stars are a good benchmark for testing the asteroseismic scaling relations. We can compare the results of our computations in the previous sections 
with the asteroseismic and non-asteroseismic properties of solar-like oscillating RGs in EBs (OEB sample).

Gaulme et al. (2016) have already confirmed discrepancies between quantities ($M$ and $R$) derived from asteroseismic and non-asteroseismic methods.
We first compare \numax~ and $\braket{\Dnu}$. In Fig. 7, the observed \numax~ is plotted \wrt $\nu_{\rm maxsca}$ where $\nu_{\rm maxsca}$
(in units of $\numax_{\odot}=3050$  \muHz)
is computed using equation (5) assuming $f_{\nu_{\rm max}}=1$, and $g$ is computed from $M_{\rm rv}$ and $R_{\rm rv}$. 
There is a small but significant difference 
between \numax~ and $\nu_{\rm maxsca}$. The fitting line is $\numax=1.038\nu_{\rm maxsca}$. This fixes that $f_{\nu_{\rm max}}=1.038$ for the OEB sample.
\begin{figure}
\includegraphics[width=101mm,angle=0]{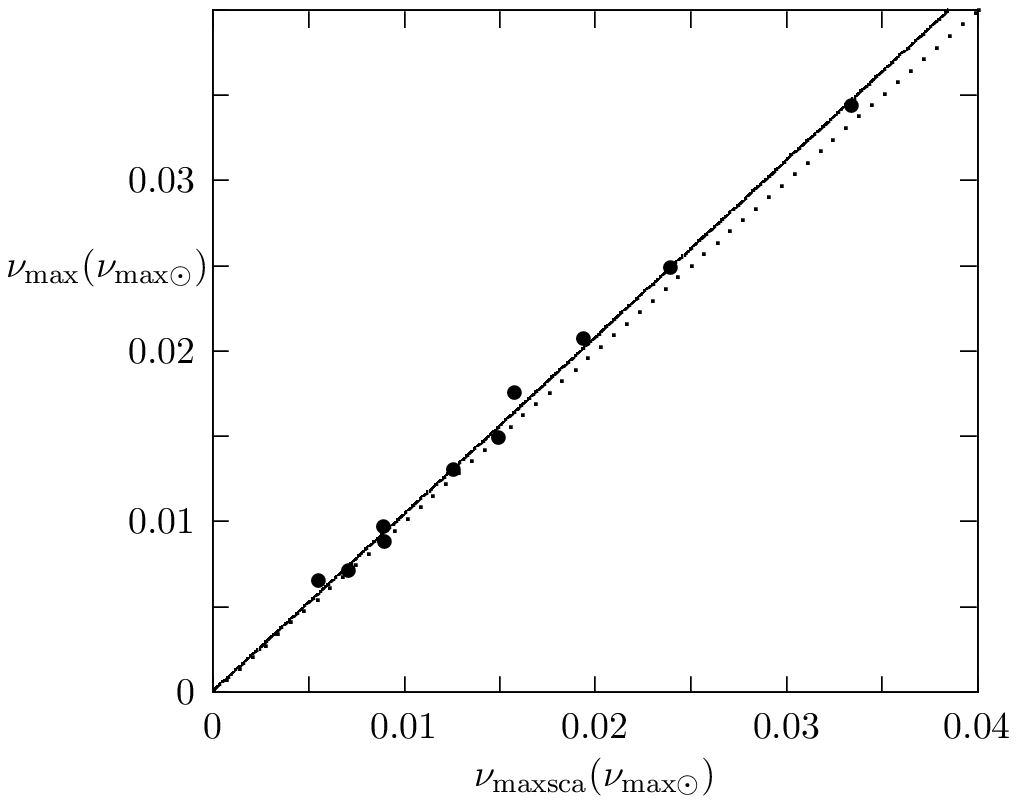}
\caption{The figure plots \numax~ of the 10 RGs in EBs (OEB sample) \wrt $\nu_{\rm maxsca}$. The solid line is for the fitting line $\numax=1.038\nu_{\rm maxsca}$, while
the dotted line represents $\numax=\nu_{\rm maxsca}$. 
}
\end{figure}

{Kallinger et al. (2018) consider OEBs in Gaulme et al. (2016), and RGs and RCs in the two open clusters, NGC 6791 and 6819. They derive non-linear scaling relations, and find that the ratio $g/\sqrt{\teff}$ is proportional to $\numax^{1.0075}$. 
The gravities from $\numax=1.038\nu_{\rm maxsca}$ and from equation (16) of Kallinger et al.  (2018) (i.e. $g_{\rm sca}$ and $g_{\rm K}$, respectively) are in very good agreement.
The mean difference between $g_{\rm sca}$  and $g_{\rm K}$ 
is about 1.1 per cent. 
While $g_{\rm sca}$ is about 0.6 per cent greater than the gravity ($g_{\rm rv}$) derived from binary dynamics, $g_{\rm K}$ is about
0.5 per cent less than $g_{\rm rv}$. The differences between the results of the two studies are mainly a result of the use of different solar data. 
For example, they take $\numax_{\odot}=3140$ \muHz.
}

\begin{figure}
\includegraphics[width=101mm,angle=0]{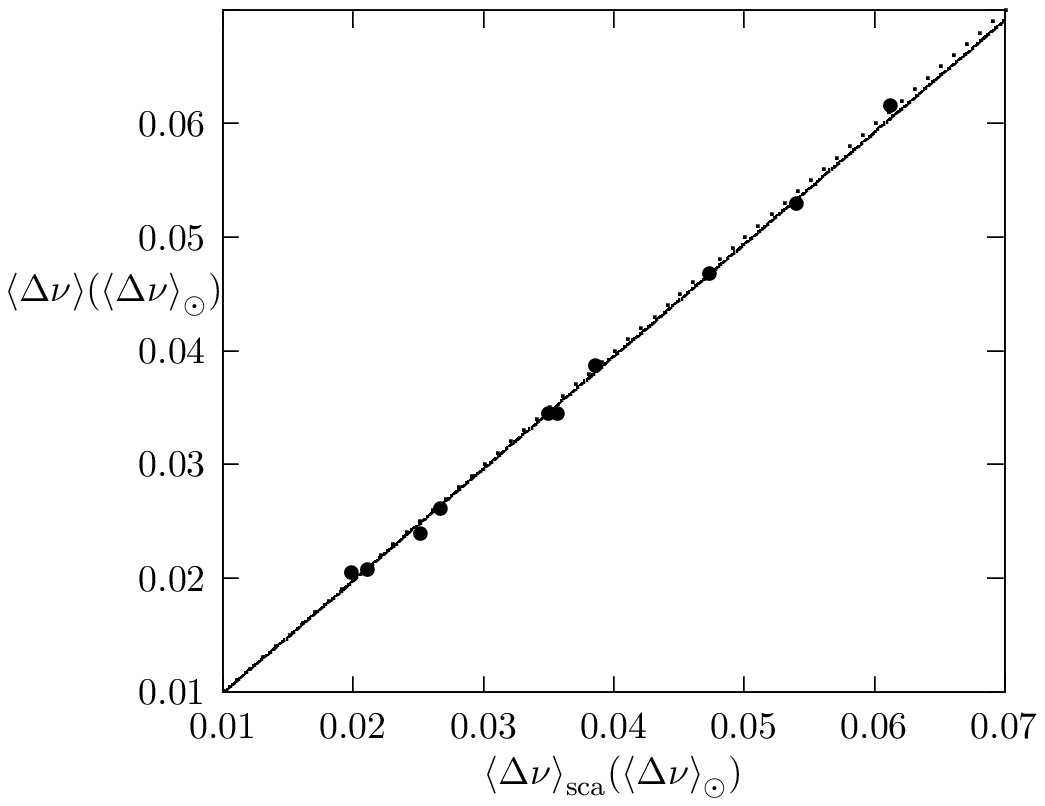}
\caption{The figure plots $\braket{\Dnu}$ of the OEB sample \wrt $\braket{\Dnu_{\rm sca}}$. The solid line shows the fitting line 
$\braket{\Dnu}=0.986\braket{\Dnu_{\rm sca}}$, while
the dotted line represents $\braket{\Dnu}=\braket{\Dnu_{\rm sca}}$. 
}
\end{figure}
In Fig. 8, $\braket{\Dnu}$ is plotted with respect to $\braket{\Dnu_{\rm sca}}$ {found from the binary masses and radii},
in solar unit ($\braket{\Dnu_{\odot}}=135.1$ \muHz).
They seem in good agreement. Nevertheless, the slope of the fitting line is {0.986. The difference, as much as 1.4 per cent,
might be considered as very small, although the influence of such a difference on $R$ and $M$ is about 3 and 6} per cent, respectively.
Therefore, we take into account this discrepancy in the scaling relations. 

{The mean densities $\rho_{\rm sca}$ from the scaling relation with $f_{\Delta \nu}=0.986$ and from equation (17) ($\rho_{\rm K}$, for RGs) of Kallinger et al. (2018) are in very good agreement.
The mean difference between $\rho_{\rm sca}$ and $\rho_{\rm K}$ is about $-1$ per cent.}

{If we plot $R_{\pi}$ from the {Y16 DR2 sample (with the BC from MIST and $A_V=0$)} and $R_{\rm rv}$ of the 10 RGs together with respect to $R_{\rm sca}$,
we obtain 
$$
R=(1.067\pm 0.013) R_{\rm sca}^{0.940\pm 0.005}.
$$
Here, $R$ is $R_{\pi}$ for the Y16 DR2 sample and $R_{\rm rv}$ for OEB sample. 
For some of the stars in the Y16 DR2 sample, $A_V$ is very large.
This prevents the two samples (Y16 DR2 and OEB) from being compatible with each other.
}

{Brogaard et al. (2018) also found the mass and radius of the stars from the binary dynamics for the three systems among the 10 OEBs. 
We only used the data of Gaulme et al. (2016) to ensure regularity in the data in our analysis.}

\subsection{Comparison of distances and radii of RGs in {APOKASC-2}}
\subsubsection{Comparison of distances of RGs using $V$}
We compute the asteroseismic and non-asteroseismic parameters of the VRG stars using the $BC$ tables of MIST.
The {\teff~ and metallicity of the RGs are taken from APOKASC-2.}
The {mean} difference between $d_{\pi}$ and $d_{\rm sca}$ is about 12.5 per cent,
much greater than the corresponding difference for the {Y16 DR2 sample}. 
{In Fig. 9, $\log R_{\pi}$ of the VRG stars (with extinction) is plotted \wrt $\log R_{\rm sca}$.
{The solid line shows the fitted line, $\log R_{\pi}=(0.949\pm 0.012) \log R_{\rm sca}+{0.083\pm 0.012}$.}
This equation can also be written as $R_{\pi}=(1.211\pm 0.032) R_{\rm sca}^{0.949\pm 0.012}$.}
This relationship between the radii for this group of stars is very similar to that for the {Y16 DR2 sample} (see Section 4.2 and Fig. 6). 
\begin{figure}
\includegraphics[width=108mm,angle=0]{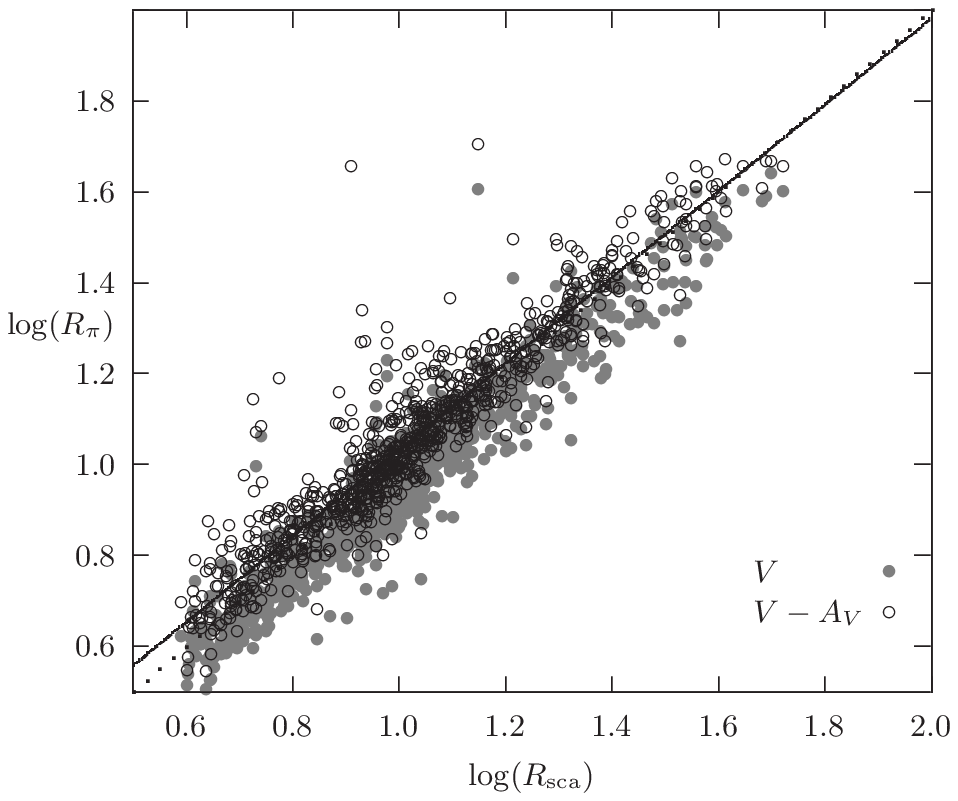}
\caption{The figure plots $\log R_{\pi}$ of {the VRG stars with and without extinction} \wrt $\log R_{\rm sca}$ in units of solar radius.
{The solid line shows the fitted line $\log R_{\pi}=(0.949\pm0.012) \log R_{\rm sca}+{0.083\pm 0.012}$}, while the dotted line denotes $\log R_{\pi}=\log R_{\rm sca}$. 
}
\end{figure}
%





The difference between $d_{\rm sca}$ and $d_{\pi}$ is greater than 50 per cent for some stars.
Such a big difference may have some unusual causes. For KIC 10273236, for example, the unusual difference between 
$d_{\rm sca}$ and $d_{\pi}$, about 62 per cent, 
seems to arise from {its} $V$. If we plot $V$ of stars  \wrt $K$, there is a strict relationship between them. Some of the 
stars do not obey this relation. KIC 10273236 is among these stars and its $V$ according to its $K$ must be about 12.25 rather than 
10.2. If we take $V$ as 12.25, the difference reduces to 3 per cent.
{$A_V$ of the VRG stars range from 0.1 to 4.0. The ratio $R_{\pi}/R_{\rm sca}$ depends on $A_V$ if $A_V>0.6$. 
If we exclude the stars with $A_V>0.6$, 
we obtain the relationship between $R_{\pi}$ and $R_{\rm sca}$ as
$$
R_{\pi}=(1.131 \pm 0.037) R_{\rm sca}^{0.967\pm 0.010}.
$$
}
{The mean difference between $\log R_{\pi}$ and $\log R_{\rm sca}$ is 0.02 (5 per cent). 
For the VRG sample without extinction, $\log R_{\pi}-\log R_{\rm sca}$= $-0.048$ (12 per cent).
These results show that the extinction improves the relationship between $R_{\pi}$ and $R_{\rm sca}$.
For the Y16 DR2 sample with the $V$-band magnitude, however, we see a better agreement when the extinction effect is excluded. The reason for this result is that
the two samples have similar range for $A_V$, but very different distance ranges. 
The Y16 DR2 sample is very close to the Sun in comparison to VRG, so $A_V$ of the Y16 DR2 sample should be smaller than that of VRG.
}
\subsubsection{Comparison of distances and radii of RGs using $G$ magnitude}

\begin{figure}
\includegraphics[width=108mm,angle=0]{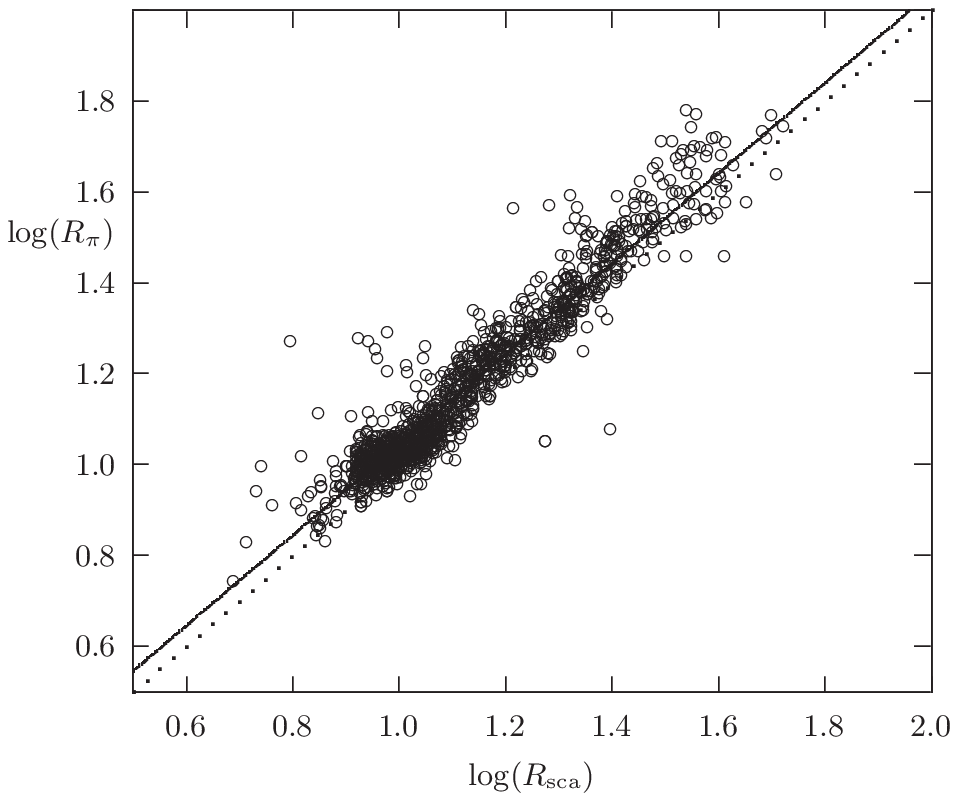}
\caption{The figure plots $\log R_{\pi}$ of {the GRG stars} with interstellar extinction \wrt $\log R_{\rm sca}$ in solar units.  
The solid line shows the fitting line: $\log R_{\pi}=(0.996\pm 0.012) \log R_{\rm sca}+{0.048\pm0.013}$. 
The dotted line denotes $\log R_{\pi}=\log R_{\rm sca}$.}
\end{figure}

The {$V$ magnitudes of stars in the SIMBAD data base are compiled from different sources with different $V$ bandwidths,
but most of them are from TYCHO-2 (H{\o}g et al. 2000). 
This may cause some problems because $BC$ depends on the bandwidth of measurements.
Therefore, we also use the {\it Gaia} catalogue and we use $G$ magnitudes for the computation of $BC$. 
In Fig. 10, $\log R_{\pi}$ of the GRG stars is plotted \wrt their $\log R_{\rm sca}$.
For this figure, \teff~ is taken from APOCASK-2 and interstellar extinction is taken into account.
The fitting curve is derived as 
$$
\log R_{\pi}=(0.996\pm 0.012) \log R_{\rm sca}+{0.048\pm0.013},
$$
where $\log R_{\pi}$ is about 12 per cent greater than $\log R_{\rm sca}$.
}




%
The $d_{\rm sca}$ and $d_{\pi}$ of RG KIC 5395942/Gaia DR2 2075037891998301312 are very different from each other. 
Therefore, we neglect this star in our analysis. The problem seems to be due to its parallax (13.09 mas).

\subsubsection{Comparison of distances and radii for the JHKRG sample}
{
In Fig. 11, 
$\log R_{\pi}$ of the JHKRG stars is plotted \wrt $\log R_{\rm sca}$ for the $J$ magnitude with extinction.
{The solid line shows the fitted line, $\log R_{\pi}=(1.0217\pm 0.0029) \log R_{\rm sca}-{0.0040\pm 0.0029}$,}
{which can also be written as $R_{\pi}=0.991 R_{\rm sca}^{1.0217}$.}
Note that $\log R_{\pi}$ and $\log R_{\rm sca}$ are in such good agreement that the fitted line and the line for $\log R_{\pi}=\log R_{\rm sca}$ 
are almost indistinguishable. A similar situation appears in other figures for $H$ and $K$.
\begin{figure}
\includegraphics[width=106mm,angle=0]{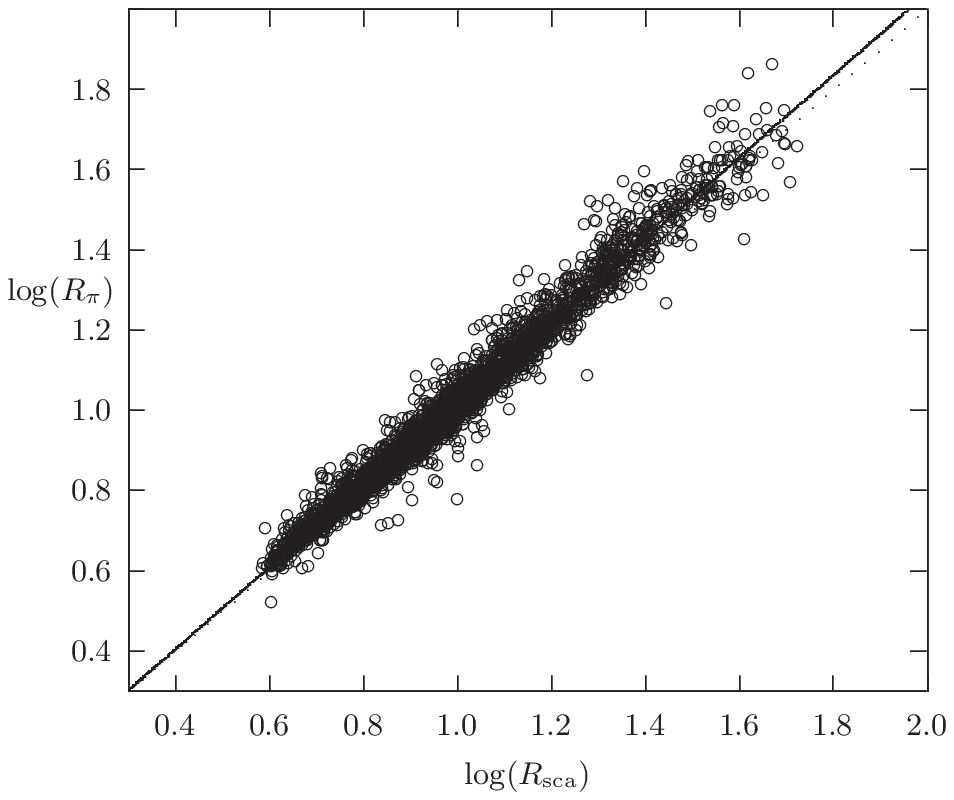}
\caption{The figure plots $\log R_{\pi}$ of {the JHKRG stars} for $J$ \wrt $\log R_{\rm sca}$ in units of solar radius.
{The solid line shows the fitted line $\log R_{\pi}=(1.0217\pm 0.0029) \log R_{\rm sca}-{0.0040\pm 0.0029}$,} while the dotted line denotes $\log R_{\pi}=\log R_{\rm sca}$. 
}
\end{figure}
}

{
Most of the JHKRG stars have a distance of less than $4000$ pc.
The distance of the JHKRG stars extends up to $10 000$ pc. However, at these distances, asteroseismic and non-asteroseismic parameters are usually not in agreement.
For most of the cases, the problem seems to be the {\it Gaia} DR2 parallax, as in the case of KIC 5395942.
}

\subsection{Comparison of masses}
\begin{figure}
\includegraphics[width=101mm,angle=0]{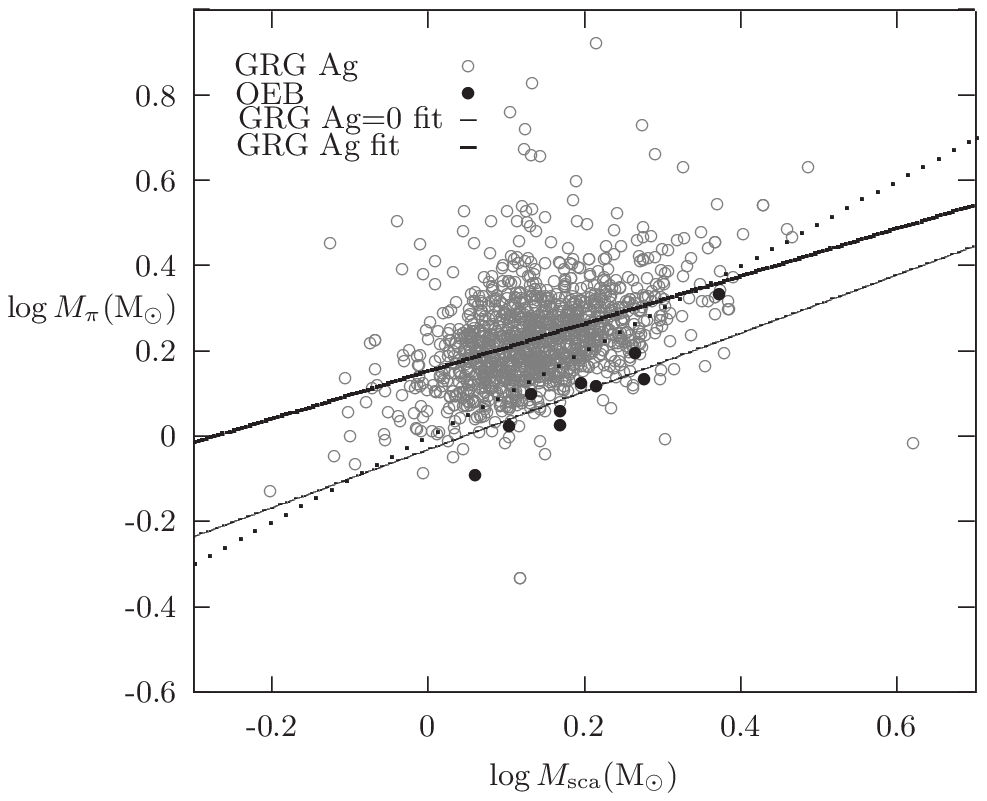}
\caption{The figure plots $M_{\pi}$ of {the GRG stars} with extinction \wrt $M_{\rm sca}$ in solar units. 
The thick solid line is the fitting curve for {the GRG stars} with $T_{\rm A}$: $\log M_{\pi}=(0.558\pm 0.034) \log M_{\rm sca}+{0.152\pm 0.006}$,
$M_{\pi}=(0.930\pm 0.053)M_{\rm sca}^{0.683\pm0.045}$. 
The thin solid line shows the fitted line for the GRG without extinction, $M_{\pi}=(0.930\pm 0.053)M_{\rm sca}^{0.683\pm0.045}$,
while the dotted line denote $\log M_{\pi}=\log M_{\rm sca}$.
The OEB sample (filled circles) is also shown.
}
\end{figure}

The same gravity $g_{\rm sca}$ and \teff~ are used in computation of  asteroseismic and non-asteroseismic distance, radius and mass. 
Therefore, $\delta g$ is equal to zero. This implies that $\delta M/M=  2\delta R/R =2(R_{\pi}-R_{\rm sca})/R_{\rm sca}$.

{
In our analysis, we use $\log g_{\rm sca}$ to compute $BC$ and therefore $R_{\pi}$ is not a purely non-asteroseismic
parameter. Nevertheless, the use of $\log g_{\rm sca}$ has a very limited effect on $R_{\pi}$ because $BC$ is not a very sensitive function
of $\log g$. However, this is not the case for $M_{\pi}$. Therefore, $M_{\pi}$ and $M_{\rm sca}$ are not independent.
}
 
Fig. 12 plots $M_{\pi}$ of {the GRG stars} without extinction \wrt $M_{\rm sca}$. The fitted curve is found to be 
$$
M_{\pi}=(0.930\pm 0.053) M_{\rm sca}^{0.683\pm 0.045}.
$$
The fitting line for the GRG stars with extinction is also plotted in Fig. 12. For both cases,
$M_{\rm sca}$ and $M_{\pi}$ are not in good agreement.


Also shown are the solar-like oscillating RGs in the OEB sample. Their positions in Fig. 12 share the same area with the most of {the GRG stars}, except for the RG component with $M_{\rm sca}$ 
of about 2.5 \MSbit.

%
%
For the OEB sample, if the power of $M_{\rm sca}$ is taken as $0.698$, then
$$
M_{\pi}=(0.926\pm 0.041) M_{\rm sca}^{0.683}.
$$
This is {close} to the curve for {GRG} without extinction.
 
{Fig. 13 plots $\log M_\pi$ of the JHKRG stars for $K$ magnitude with respect to $\log M_{\rm sca}$ in solar units.
The fitting line for the JHKRG stars with $\Delta \pi/\pi<0.02$ is {obtained as
$$\log M_{\pi}=(0.906\pm 0.030)\log M_{\rm sca}+0.000\pm 0.004,$$}
which can also be written as $M_{\pi}=M_{\rm sca}^{0.906}$.
For JHKRG, $M_{\rm sca}$ and $M_{\pi}$ are in good agreement (see also Section 5.4 and Fig. 19).
}
\begin{figure}
\includegraphics[width=101mm,angle=0]{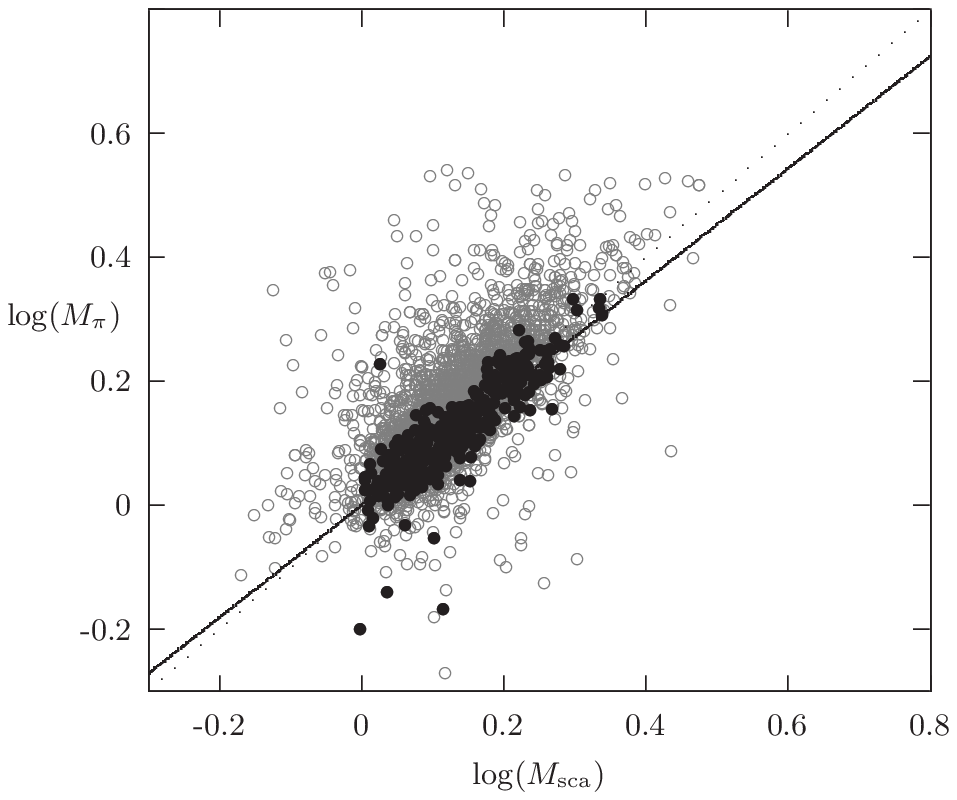}
\caption{The figure plots $\log M_{\pi}$ of {the JHKRG stars} (circles) for the $K$-band magnitude \wrt $\log M_{\rm sca}$ in solar units. 
The filled circles represent the stars with $\Delta \pi/\pi<0.02$.
{The solid line shows the fitting line for them, $\log M_{\pi}=(0.906\pm 0.030)\log M_{\rm sca}+0.000\pm 0.004$, and}
the dashed line denotes $\log M_{\pi}=\log M_{\rm sca}$. 
}
\end{figure}







\subsection{${f_{\Delta \nu}}$ and ${f_{\nu_{\rm max}}}$ from comparison of masses and radii}
\begin{figure}
\includegraphics[width=101mm,angle=0]{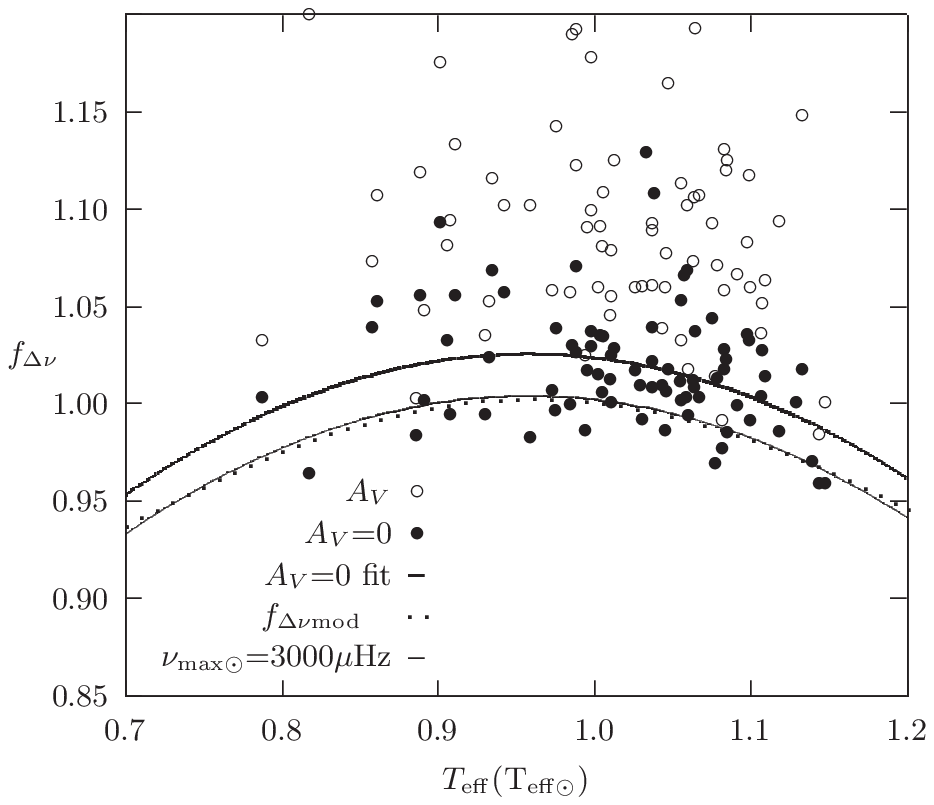}
\includegraphics[width=101mm,angle=0]{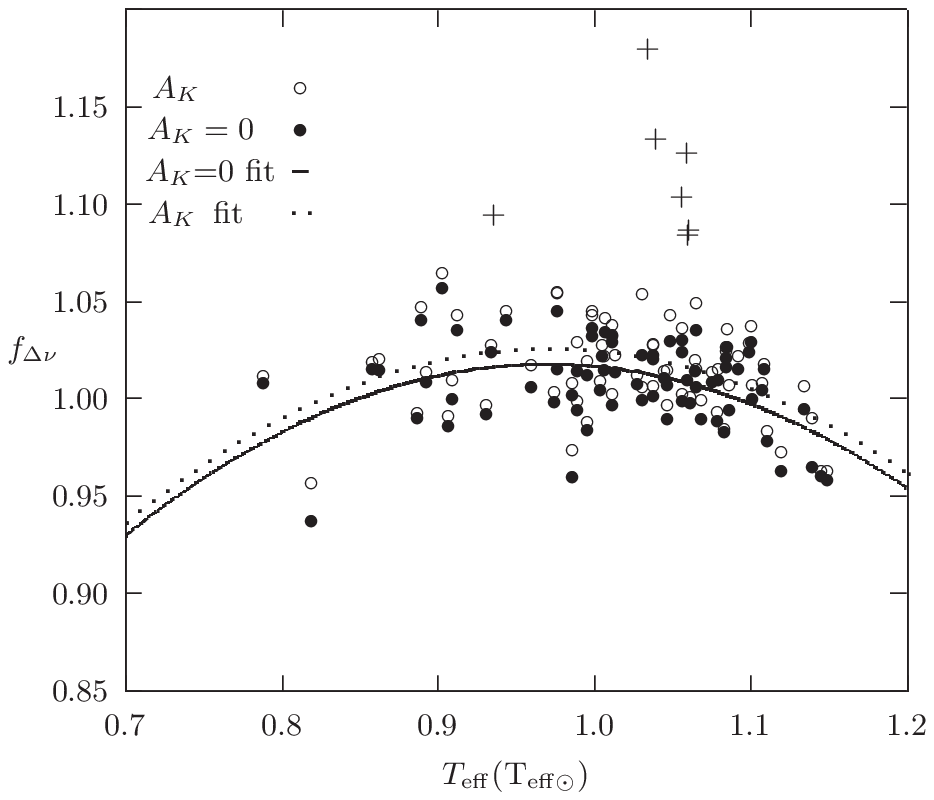}
\caption{(a) For the {Y16 DR2 sample} with (circles) and without (filled circles) interstellar extinction for the $V$-band magnitude, $f_{\Delta \nu}$ is plotted \wrt $\teff$ in solar units. 
The thick solid line shows the fitting curve obtained by excluding three stars (open circles) in the upper part of the figure, 
$f_{\Delta \nu}=(-1.280\pm 0.433)( T_{\rm eff}/{\rm T}_{\rm eff \odot}-0.958)^2+1.031\pm 0.005$.     
The dotted line shows $f_{\Delta \nu}$ derived from the interior models by Y\i ld\i z et al. (2016). 
If we take $\nu_{\rm max\odot}$= 3000$ \mu$Hz, $f_{\Delta \nu}$ for $A_V=0$ is in very good agreement with that of  Y\i ld\i z et al. (2016).
(b) $f_{\Delta \nu}$ for the $K$-band magnitude with and without extinction, $f_{\Delta \nu}=(-1.219\pm 0.313)( T_{\rm eff}/{\rm T}_{\rm eff \odot}-0.970)^2+1.026\pm 0.004$ and $f_{\Delta \nu}=(-1.207\pm 0.301)( T_{\rm eff}/{\rm T}_{\rm eff \odot}-0.970)^2+1.018\pm 0.003$, respectively. The fitting lines for these two cases are very similar to each other.
}
\end{figure}

In principle, we can obtain expressions for $f_{\nu_{\rm max}}$ and $f_{\Delta \nu}$ from the ratios $M_{\pi}/M'_{\rm sca}$ and $R_{\pi}/R'_{\rm sca}$. 
{Assuming $M_{\pi}$ and $R_{\pi}$ as exact values and using equations (7) and (8),
we obtain 
\begin{equation}
\frac{R_{\rm \pi}}{R_{\rm sca}}=
\frac{f_{\Delta \nu}^2}{f_{\nu_{\rm max}}},
\end{equation}
and 
\begin{equation}
\frac{M_{\rm \pi}}{M_{\rm sca}}=
\frac{f_{\Delta \nu}^4}{f_{\nu_{\rm max}}^3}.
\end{equation}
If we take the cube of equation (9) and divide it by equation (10), the right-hand side becomes $f_{\Delta \nu}^2$. Then, we obtain
$$f_{\Delta \nu}=\sqrt{(R_{\pi}/R_{\rm sca})^3/(M_{\pi}/M_{\rm sca})}.$$
Similarly, we derive the expression for $f_{\nu_{\rm max}}$ as 
$(R_{\pi}/R_{\rm sca})^2/(M_{\pi}/M_{\rm sca})$.
}
%

{
This formulation would be valid if we precisely determined $\log g_{\rm S}$ from stellar spectra. Unfortunately, $\log g_{\rm S}$ is not so accurate and we have to use
$\log g_{\rm sca}$. Then, $M_{\pi}$ and $R_{\pi}$ are not purely non-asteroseismic quantities. Therefore, $f_{\nu_{\rm max}}$ and $f_{\Delta \nu}$ are inter-related (see Section 5).
}

For MS and SG stars, $f_{\Delta \nu}$ is a function of \teff~ (or $\Gamma_{\rm 1s}$) as in Y\i ld\i z et al. (2016).
{However, this relation is derived for the solar composition and $f_{\Delta \nu}$ may also be a function of metallicity, for example. 
Fig. 14(a) plots $f_{\Delta \nu}$ of the {Y16 DR2 sample }\wrt \teff~, assuming $f_{\nu_{\rm max}}=1$. $BC$ is computed from the MIST tables.} Three stars (KIC 8379927, 
KIC 9025370 and KIC 11772920) are taking place in the upper
part of Fig. 14(a). The fitting curve is obtained by neglecting these three stars as 
\begin{equation}
f_{\Delta \nu}=(-1.280\pm 0.433)( T_{\rm eff}/{\rm T}_{\rm eff \odot}-0.958)^2+1.031\pm 0.005. \nonumber    
\end{equation}


{Fig. 14(a) also shows $f_{\Delta \nu}$ obtained from the interior models by Y\i ld\i z et al. (2016)
due to a variation of $\Gamma_{\rm 1s}$. The difference between these two expressions for $f_{\Delta \nu}$ is almost constant and is about 0.02. 
Such a small difference may be the result of the modelling of the outer regions
or the uncertainties in any of the observed quantities such as \teff~  and $\pi$.
Another way of removing the discrepancy is to use $\nu_{\rm max\odot}$=3000$ \mu$Hz.}
{If we compute $BC$ from Lejeune et al. (1998) rather than MIST and use  $V$ magnitude, we obtain} 
$f_{\Delta \nu}=-1.286(T_{\rm eff}/{\rm T}_{\rm eff \odot}-0.96)^2+1.017$. 


In Fig. 14(b), $f_{\Delta \nu}$ is plotted for the $K$-band magnitude with and without extinction. For the extinction case,
we obtain
\begin{equation}
f_{\Delta \nu}=(-1.219\pm 0.313)( T_{\rm eff}/{\rm T}_{\rm eff \odot}-0.970)^2+1.026\pm 0.004. 
\end{equation}


We can use spectroscopically derived  $\log g_{\rm S}$ in the computation of $M_{\pi}$ of the {Y16 DR2 sample}, and we try to obtain $f_{\nu_{\rm max}}$. However,
the accuracy of $\log g_{\rm S}$ does not allow us to reach a definite relationship for $f_{\nu_{\rm max}}$, 
{because, the spectroscopic $\log g_{\rm S}$ is calibrated on asteroseismic $\log g_{\rm sca}$.} 


Similarly, we can also derive an expression for $f_{\Delta \nu}$ of RGs {as a function of \teff},
{again assuming $f_{\nu_{\rm max}}=1$}. In Fig. 15, $f_{\Delta \nu}$ is plotted \wrt $\teff/\teff_{\odot}$ for {the GRG stars}.
We notice that the most of the stars have $f_{\Delta \nu}$ significantly less than 1 for the case without extinction. 
{The fitting curve is obtained as 
\begin{equation}
f_{\Delta \nu}=(-3.28\pm 1.12 )( T_{\rm eff}/{\rm T}_{\rm eff \odot}-0.777\pm 0.007)^2+0.964\pm 0.002.
\end{equation}
Its maximum value is 0.964. This implies that the corrections for  $R_{\rm sca}$ and $M_{\rm sca}$ are about 7 and 14 per cent, respectively. 
Also shown in Fig. 15 is the fitting line for the $G$-band magnitude with extinction, which is very different from  equation (12).}
\begin{figure}
\includegraphics[width=101mm,angle=0]{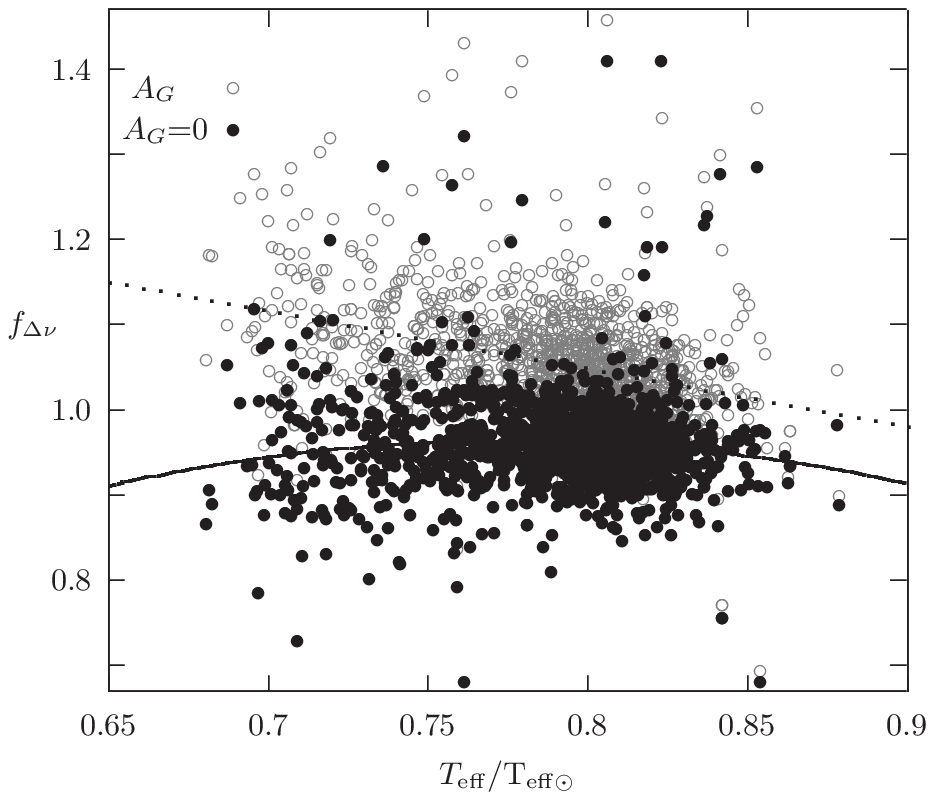}
\caption{The figure plotd $f_{\Delta \nu}$ of the GRG stars with (circles) and without (filled circles) interstellar extinction \wrt $\teff$ in solar units.
The thick solid line shows the fitting curve for $A_G=0$: $f_{\Delta \nu}=(-3.28\pm 1.12 )( T_{\rm eff}/{\rm T}_{\rm eff \odot}-0.777\pm 0.007)^2+0.964\pm 0.002$. 
The dotted line denotes the case with $A_G$: $f_{\Delta \nu}= (-0.679\pm0.053)T_{\rm eff}/{\rm T}_{\rm eff \odot} + 1.590 \pm 0.042$.
}
\end{figure}

{
There are two very important uncertainties for the expression for $f_{\Delta \nu}$ given in equation (12).
These uncertainties are due to the assumption about $f_{\nu_{\rm max}}=1$ and the $G$ magnitude. 
The use of $J$, $H$ and $K$ magnitudes may change the situation (see Section 5.3).
}

\begin{figure}
\includegraphics[width=101mm,angle=0]{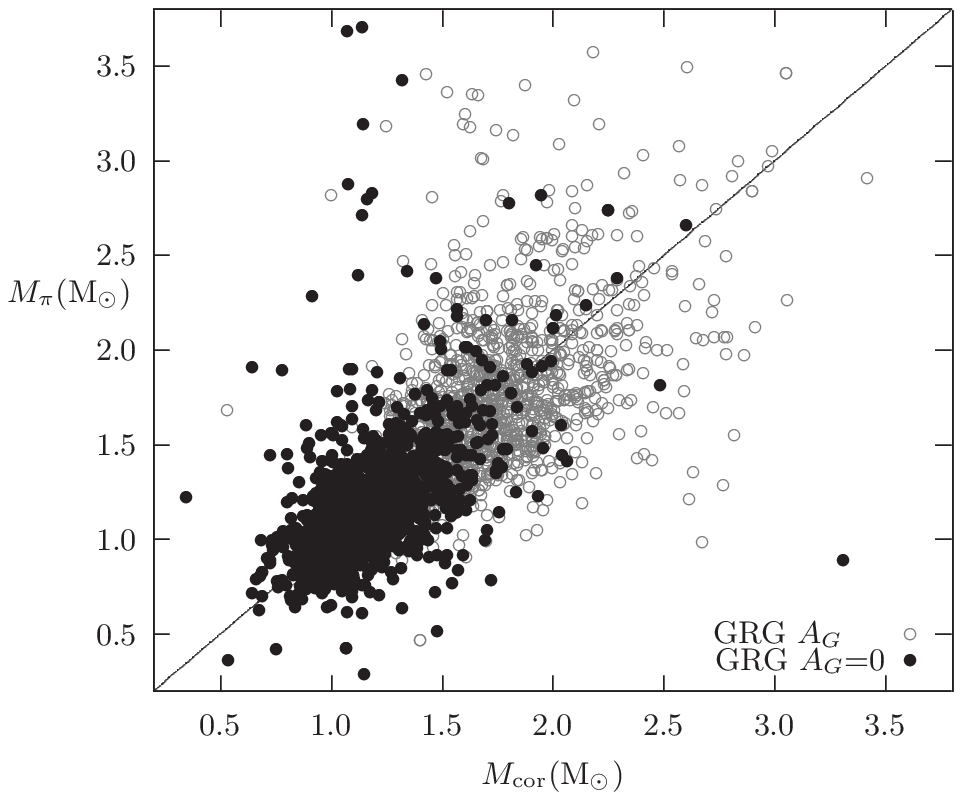}
\caption{The figure plots $M_{\pi}$ of {the GRG stars}  with (circles) and without extinction (filled circles) \wrt the corrected asteroseismic mass $M_{\rm cor}$ in solar units using 
$f_{\Delta \nu}$ derived from Fig. 15. 
The line denotes  $M_{\pi}=M_{\rm cor}$.
}
\end{figure}
In Fig. 16, $M_{\pi}$ is plotted \wrt the corrected $M_{\rm sca}$ ($M_{\rm cor}$) according to $f_{\Delta \nu}$ derived for {GRG} from Fig. 15 
(equation 12). 
The relationship between $M_{\pi}$ and $M_{\rm cor}$ is linear and the positions of the most of the stars are close to the $M_{\pi}$=$M_{\rm cor}$ line.
Most of the stars have mass in the range of 0.9-1.5 \MS.
However, there is also significant scattering. For 319 of {the GRG stars}, the difference between $M_{\pi}$ and $M_{\rm cor}$ is less 
than 5 per cent. There are some stars with mass $0.5 <M<0.9 \MS$. The MS lifetime of these stars
is greater than the age of the Galaxy. Therefore, they cannot be in the RG branch as long as these masses are their initial masses.
However, at least for some of these stars, both $M_{\rm cor}$ and $M_{\pi}$ are small. 
This leads us to rely on the computed masses. Then, these stars might have experienced serious mass loss during their past evolution.

{If we use $f_{\Delta \nu}$ derived from Fig. 15 for the case with extinction, we obtain different values for the mass of the GRG stars. The mean mass difference 
between the cases with and without extinction is about 0.55 \MS. }

{
The $f_{\Delta \nu}$ for the VRG stars is completely different from that given in Fig. 15 (dotted line). It has a positive slope.
This conflicting situation may also be due to extinctions for $V$- and $G$-band magnitudes.}

\subsection{Effective temperature tests }
{In this section, we want to test the effect of the uncertainty of \teff~ on the differences between asteroseismic ($d_{\rm sca}$, $R_{\rm sca}$ and $M_{\rm sca}$) and non-asteroseismic quantities ($d_{\pi}$, $R_{\pi}$ and $M_{\pi}$).
We try to make the corrected $R_{\rm sca}$ and $R_{\pi}$ equal by changing only \teff.}
We first clarify how $R_{\rm sca}$ and $R_{\pi}$ are sensitive to changes in \teff.
For $R_{\rm sca}$, we find from the scaling relation that
\begin{equation}
\frac{\Delta R_{\rm sca}}{R_{\rm sca}}=\frac{1}{2}\frac{\Delta T_{\rm eff}}{T_{\rm eff}}.
\end{equation}
For $R_{\pi}$, the situation is slighty more complicated. 
We first compute $L_\pi$ from the absolute magnitude and $BC$. The change in \teff~ influences $BC$ and hence $L_\pi$.
{If we neglect the temperature dependence of $BC$,}
then the dependence of 
$R_{\pi}$ is determined by $R_{\pi}=(L_\pi/4\pi \sigma)^{1/2}/\teff^2$ (for the coolest stars, this is not the case; see below). 
For a given value of $L_\pi$, we can show that
\begin{equation}
\frac{\Delta R_{{\pi}}}{R_{{\pi}}}={-2}\frac{\Delta T_{\rm eff}}{T_{\rm eff}}.
\end{equation}
While $R_{\rm sca}$ is proportional to $\teff^{1/2}$, $R_{\pi}$ is inversely proportional to $\teff^{2}$.
As the \teff~ dependences of $R_{\rm sca}$ and $R_{\pi}$ are completely different, we attempt to remove the 
difference between the asteroseismic and non-asteroseismic quantities
by changing only \teff.

The estimated \teff~ ($T'_{\rm eff}$) as a function of $R_{\rm sca}$ and $R_{\pi}$ is found from equations (13) and (14) as
\begin{equation}
T'_{\rm eff}=\teff+\Delta \teff=\teff(1+x_T)
\end{equation}
where
\begin{equation}
x_T=\frac{\Delta \teff}{\teff}=\frac{R_{\pi}-R_{\rm sca}}{2R_{{\pi}}+R_{\rm sca}/2}.
\end{equation}
If $R_{\pi}$ is less than $R_{\rm sca}$, then $T'_{\rm eff}$ is lower than \teff; otherwise, $T'_{\rm eff}$ is higher than \teff.

$BC$ is almost constant for about $5000<\teff<6000$ K. For $\teff<4100$ K, however, $BC$ is a very sensitive function of 
\teff. Therefore, we follow an iterative method to find the most appropriate $T'_{\rm eff}$ for a star for which $R_{\pi}=R_{\rm sca}$. 
We first compute all the asteroseismic and non-asteroseismic quantities using the catalogue value of \teff~ and then we
find $T'_{\rm eff}$. We recompute all the quantities using the new effective temperature and we repeat the procedure until the effective 
temperature is fixed at a certain value.  

However, for $\teff <$ 4100 K, where $\partial BC/\partial \teff$ is not small, we find that the coefficient ($-2$) on the right-hand side of 
{equation (14) becomes $-2-(\partial BC/\partial \ln \teff)\ln (10)/5$. For $\teff <$ 4100 K, we obtain 
$(\partial BC/\partial \ln \teff)=3.934$ from the MIST tables.
This method can be applied to the much more precise data for parallax than {\it Gaia} DR2.
}

The difference between $d_{\rm sca}$ and $d_{\pi}$ can be removed in a variety of ways, one of which is to change \teff. 
The temperature difference between $T_{\rm A}$ and effective temperature from LAMOST (Zong et al. 2018) reaches 120 K. If we 
decrease $T_{\rm A}$ by about 183 K, there is no mean difference between $d_{\rm sca}$ and $d_{\pi}$. 
{However, a change of 183 K in the APOGEE temperature scale would be very large.
}

\section{Parallax offset and scaling relations}

The difference between asteroseismic and non-asteroseismic distance can arise either from the difference between true ($\pi_{\rm true}$) and {\it Gaia} DR2 parallaxes or from the
difference between true radius ($R_{\rm true}$) and $R_{\rm sca}$. If there is an offset, then $\Delta \pi=\pi-\pi_{\rm true}$ is a constant. $R_{\rm true}$ can be written as a function of $f_{\nu_{\rm max}}$ and $f_{\Delta \nu}$ :
$R_{\rm true}=R_{\rm sca} f_{\Delta \nu}^2/f_{\nu_{\rm max}}$. If $R_{\rm sca}$ is perfect, then $f_{\Delta \nu}=f_{\nu_{\rm max}}=1$, so the defect is pertaining to the {\it Gaia} DR2 data and we can obtain $\Delta \pi$. 
If $\pi$ is perfect, then $\Delta \pi=0$ and we can at least find $f_{\Delta \nu}^2/f_{\nu_{\rm max}}$ from the relation between $R_{\rm true}$ and $R_{\rm sca}$ (see below). 
However, we can do something better than these two options, 
if we can obtain a relationship between $f_{\Delta \nu}^2$ , $f_{\nu_{\rm max}}$ and $\Delta \pi$ {(see also Khan et al. 2019; Zinn et al. 2019b)}.   

The radius and distance of a star are {related through} its luminosity. We assume that $\teff$, $V$ and $BC$ are given.
If we determine the radius (e.g. from asteroseismic scaling relations), then we can find the distance from the comparison of the flux 
at the surface of the star and the flux we receive. If we know the radius, we can determine distance, or vice versa.
If the difference between 'true' and estimated radii is, for example, 2 per cent, then the difference between 
true and estimated distances is also 2 per cent {(see section 4.2)}. 

This is the case for both sets of quantities based on {\it Gaia} DR2 parallax and asteroseismic scaling relations. 
\begin{equation}
R_{\rm true}=R_{\pi} \frac{d_{\rm true}}{d_{\pi}}=\frac{R_{\pi}}{(1-\Delta \pi d_{\pi})}.  \nonumber
\end{equation}
Using the relation between $R_{\rm sca}$ and $R_{\rm true}$,
\begin{equation}
R_{\rm true} = R_{\rm sca} \frac{f_{\Delta \nu}^2}{f_{\nu_{\rm max}}} = \frac{R_{\pi}}{(1-\Delta \pi d_{\pi})}.  \nonumber
\end{equation}
Then,
\begin{equation}
f_{\nu_{\rm max}}=f_{\Delta \nu}^2 (1-\Delta \pi d_{\pi}) \frac{R_{\rm sca}}{R_{\pi}}.
\end{equation}

For MS and RG stars we know $f_{\Delta \nu}$ from models. Then, we have a relationship between $f_{\nu_{\rm max}}$ and $\Delta \pi$ from equation (17).

%
 
\subsection{Parallax offset for the MS and SG stars}
{
If we take $f_{\Delta \nu}=f_{\nu_{\rm max}}=1$, the parallax offset for the $K$-band magnitude of the Y16 DR2 sample with and without 
interstellar extinction is $-0.189$ and $-0.094$ mas, respectively. We have already derived an expression for $f_{\Delta \nu}$
for the former case (equation 11). Use of this expression gives a negligibly small offset, $\Delta \pi=0.005$ mas.
}

\subsection{Parallax offset for RG stars for approximate values of $f_{\Delta \nu}$ and $f_{\nu_{\rm max}}$}
{We have already derived $f_{\Delta \nu}$ assuming no parallax offset. If we take $f_{\Delta \nu}=f_{\nu_{\rm max}} =1$, then we can find the value of $\Delta \pi$.
For $J$-, $H$- and $K$-band magnitudes, we obtain $\Delta \pi$ as $-0.0232$, $-0.0127$ and $-0.0145$ mas, respectively.
Another way to compute $\Delta \pi$ is to use values of $f_{\Delta \nu}$ and $f_{\nu_{\rm max}}$ from the OEB sample. In that case,
$\Delta \pi$ is equal to  $-0.0621$, $-0.0506$ and $-0.0525$ mas for the $J$-, $H$- and $K$-band magnitudes, respectively.
These values of $\Delta \pi$ are in good agreement with the values given in the literature. 
In some studies on the analysis of {\it Gaia} DR2 data, potential color-, magnitude-, and spatial-dependent terms to $\Delta \pi$ are found (see, e.g., Leung \& Bovy 2019; Zinn et al. 2019a).
Similar associations can be made for our results. However, there are some cases where these dependences disappear (see below).
}
\subsection{Relationship  between parallax offset and $f_{\nu_{\rm max}}$ for RG stars with $f_{\Delta \nu}$ from models}
{
Model values of $f_{\Delta \nu}$ ($f_{\Delta \nu{\rm mod}}$) are available from Sharma et al. (2016).
$f_{\Delta \nu{\rm mod}}$ is computed using the code given by Sharma et al. (2016), as a function of \teff, $\log Z$, $M$ and $\log g$. It is in very good agreement with the
expression derived by \yildiz et al. (2016) for MS models. For RGs, however, $f_{\Delta \nu{\rm mod}}$ is very different from that given 
in Fig. 15 and its mean value for the JHKRG stars is about 0.975. This value is in good agreement with the value found from the OEB sample. 
}

{
}

{
If we use $f_{\Delta \nu{\rm mod}}$, there are now two unknowns in our analysis: 
$\Delta \pi$ and $f_{\nu_{\rm max}}$. In this subsection, we first consider $K$-band magnitude of the JHKRG stars.
If we take $f_{\nu_{\rm max}}=1$, then $\Delta \pi$ becomes about -0.044 mas. For $f_{\nu_{\rm max}}=1.038$,
the value for the OEB sample, $\Delta \pi$ is about -0.067 mas. In Fig. 17, $f_{\nu_{\rm max}}$ is plotted with respect to 
$\Delta \pi$. 
For each pair of solutions in Fig. 17, the difference ($\delta \pi$) between $\pi'=\pi-\Delta \pi$ and $\pi_{\rm sca}$ shows $\pi$ dependence.
We make a line equation fit to the graph of $\delta \pi$ and $\pi$: $\delta \pi=a\pi+b$, where
$a$ and $b$ are plotted with respect to $\Delta \pi$ in Fig. 18. 
While the slope $a$ is less than zero for $\Delta \pi=-0.067$ mas, it is greater than zero for $\Delta \pi=-0.021$, for example.
This means that at an intermediate value of $\Delta \pi$, the slope becomes zero. At $\Delta \pi=-0.0463$ mas, $f_{\nu_{\rm max}}=1.003$. 
At this point, $a$ and $b$ intersect each other and 
they are very close to zero. 
This is reasonable because the real mean difference between the two parallaxes is zero where the slope is zero.
}

{
We apply the same method to $J$ and $H$ data. A summary of the results is provided in Table 1.
Values of $\Delta \pi$ for $H$- and $K$-band magnitudes are very close to each other.
This is the case also for the values of $f_{\nu_{\rm max}}$.
}
\begin{figure}
\includegraphics[width=101mm,angle=0]{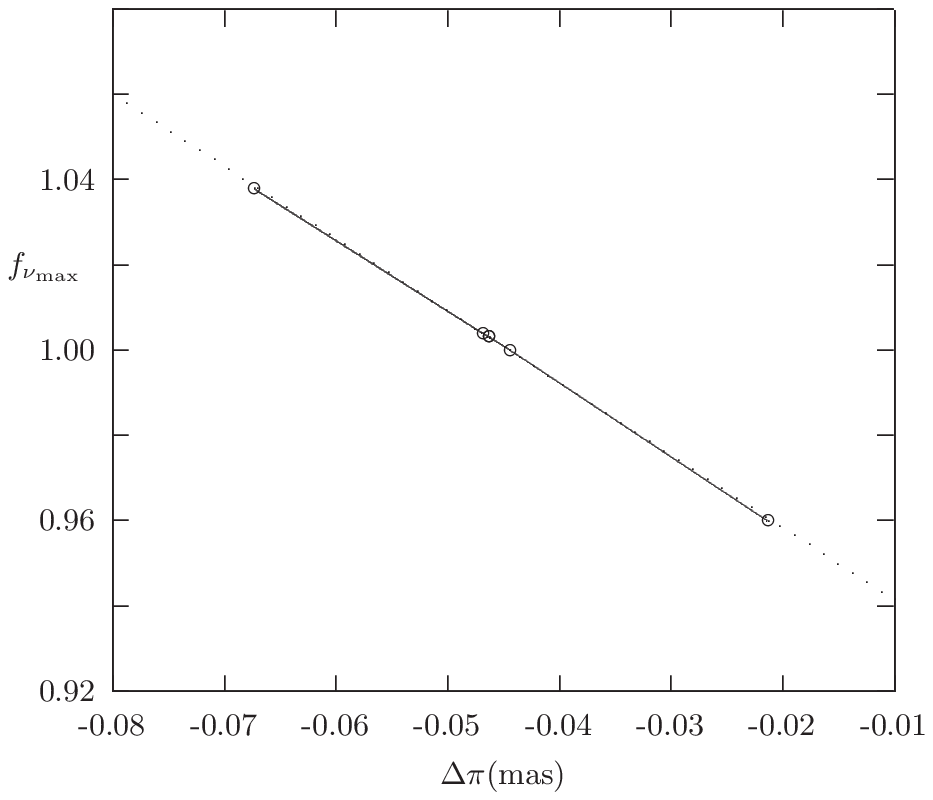}
\caption{The figure plots $f_{\nu_{\rm max}}$ \wrt $\Delta \pi$ for {JHKRG} stars with $f_{\Delta \nu{\rm mod}}$ from Sharma et al. (2016).
{This figure is for the $K$-band. Very similar figures are obtained for the $J$- and $H$-band.}
}
\end{figure}
\begin{figure}
\includegraphics[width=101mm,angle=0]{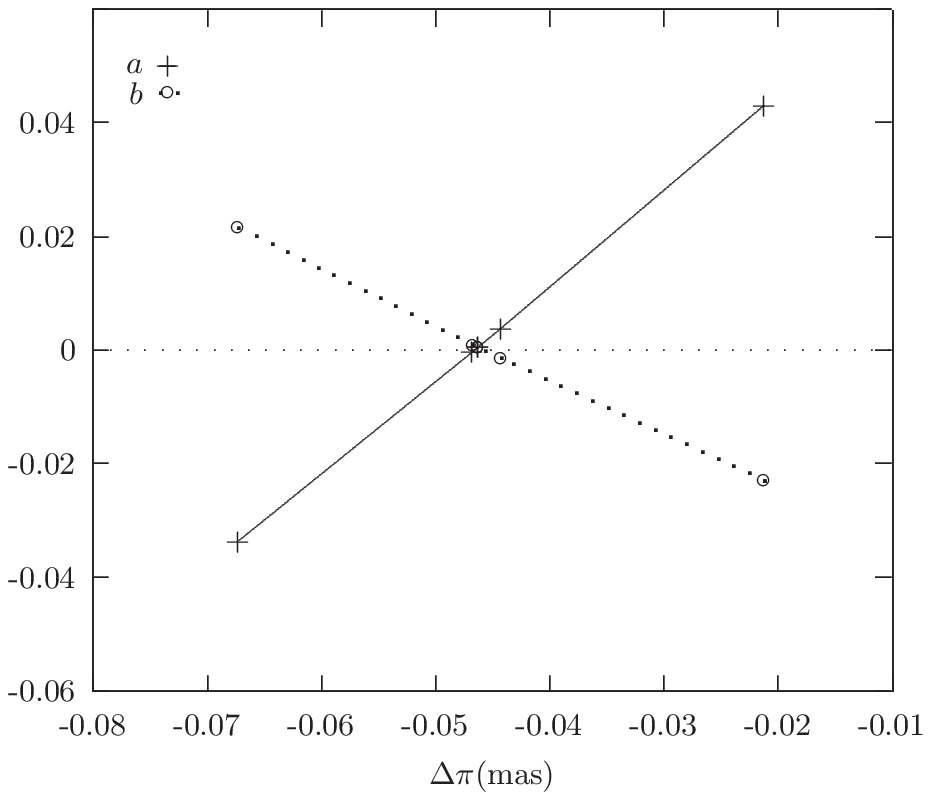}
\caption{$\delta \pi=\pi'-\pi_{\rm sca}$ is a function of $\pi$ for given values of $f_{\nu_{\rm max}}$ and $\Delta \pi$ and 
can be represented by a line equation: $\delta \pi=a \pi+b$. Here, $a$ (plus symbols) and $b$ (circles) for the $K$-band are plotted with respect to $\Delta \pi$. 
At about $\Delta \pi=-0.0463\pm0.0007$ ($f_{\nu_{\rm max}}=1.003\pm0.001$), $a$ and $b$ both vanish and intersect each other. 
The typical uncertainty of $f_{\nu_{\rm max}}$ is obtained from the uncertainties of $a$ and $b$ at $\Delta \pi=-0.0463\pm0.0007$:
$\Delta f_{\nu_{\rm max}} = \Delta a \Delta \pi + \Delta b$. The typical uncertainty for $\Delta \pi$ is found to be 0.0007 mas from the $\Delta \pi$ difference between 
the points $a=0$ and $b=0$.
}
\end{figure}
{
\begin{table}
\caption{Mean values of asteroseismic scaling parameters and parallax offset for $J$-, $H$- and $K$-band magnitudes for JHKRG.
$\overline{f_{\Delta \nu{\rm mod}}}$ is the mean of $f_{\Delta \nu{\rm mod}}$ computed from Sharma et al. (2016). 
For comparison, the scaling parameters of OEBs are also given. We notice that there is a very good agreement between the
scaling parameters for $H$ and OEBs. $\Delta M$ given in the last column is the mean difference between $M'_{\rm sca}$ and $M_{\rm AP}$.
In the last row, uncertainties in $f_{\nu_{\rm max}}$ and $\overline{f_{\Delta \nu}}$ are given for OEBs.
}
\centering
\small\addtolength{\tabcolsep}{-3pt}
\begin{tabular}{lcccr}
\hline
Data         &  $\Delta \pi$(mas) & $f_{\nu_{\rm max}}$ & $\overline{f_{\Delta \nu{\rm mod}}}$ & $\Delta M$(\MS)\\
\hline
$J$          &  $-0.0333$     &  0.966 &  0.975    &  0.148\\             
$H$          &  $-0.0424$     &  1.000 &  0.975    &  0.010\\             
$K$          &  $-0.0463$     &  1.003 &  0.975    &  $-0.001$\\             
OEBs         &   ---        &  1.038 &  0.986    &    ---\\             
$\epsilon$(OEB)            &   ---        &  0.010 &  0.006    &    ---\\             
\hline
\end{tabular}
\end{table}
}

{
We also obtain solutions for $J$-, $H$- and $K$-band magnitudes without extinction. The results are summarized 
in Table 2.
It is worth noting that the results for $H$ are very consistent with the data of OEBs.
\begin{table}
\caption{The same as Table 1 but without interstellar extinction.
}
\centering
\small\addtolength{\tabcolsep}{-3pt}
\begin{tabular}{lcccc}
\hline
Data         &  $\Delta \pi$(mas) & $f_{\nu_{\rm max}}$ & $\overline{f_{\Delta \nu{\rm mod}}}$ & $\Delta M$(\MS)\\
\hline
$J$          &  $-0.0351$     &  1.019 &  0.975    &  0.048\\             
$H$          &  $-0.0432$     &  1.036 &  0.975    &  0.103\\             
$K$          &  $-0.0459$     &  1.025 &  0.975    &  0.065\\             
OEBs         &   ---        &  1.038 &  0.986    &    ---\\             
             &   ---        &  0.010 &  0.006    &    ---\\             
\hline
\end{tabular}
\end{table}
}
\subsection{Masses of RGs}
\begin{figure}
\includegraphics[width=101mm,angle=0]{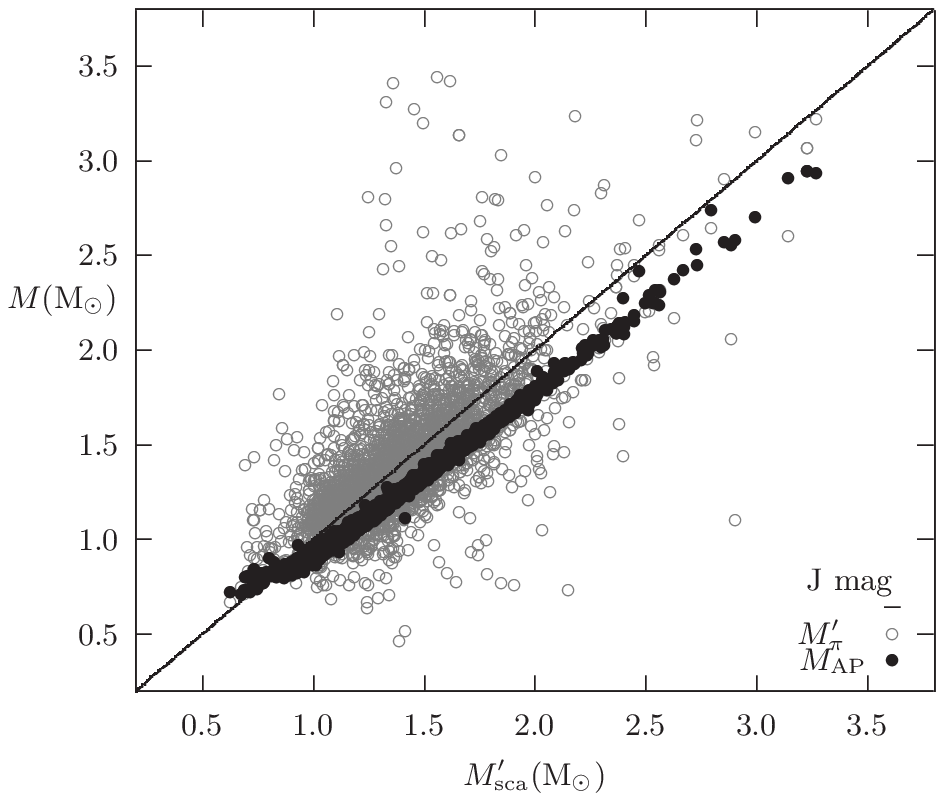}
\includegraphics[width=101mm,angle=0]{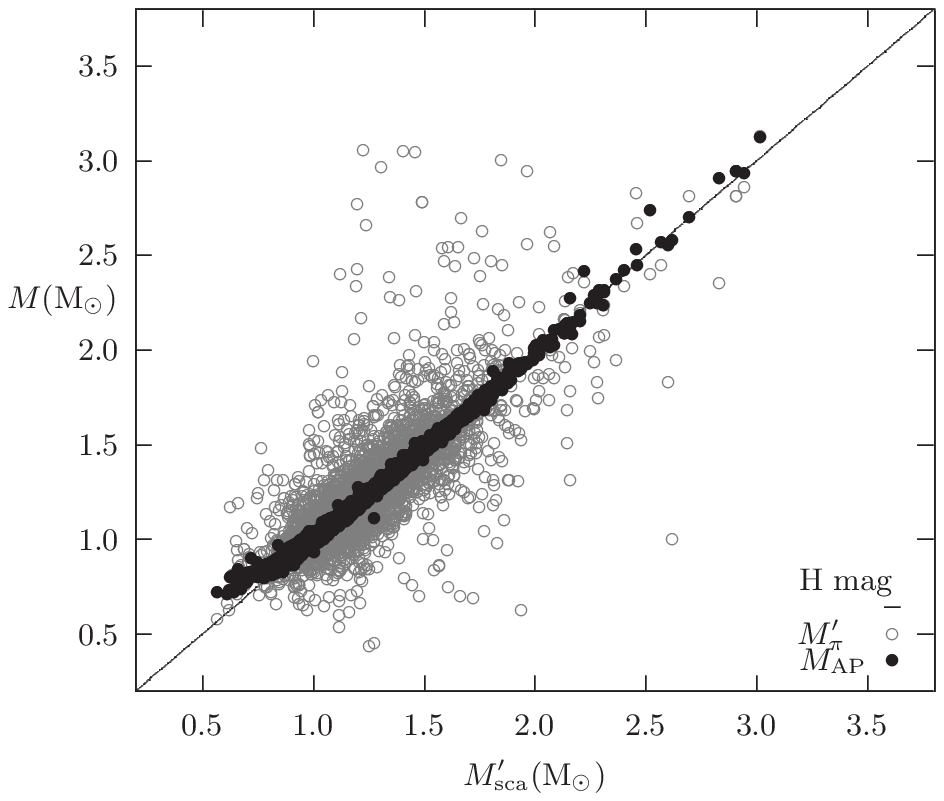}
\includegraphics[width=101mm,angle=0]{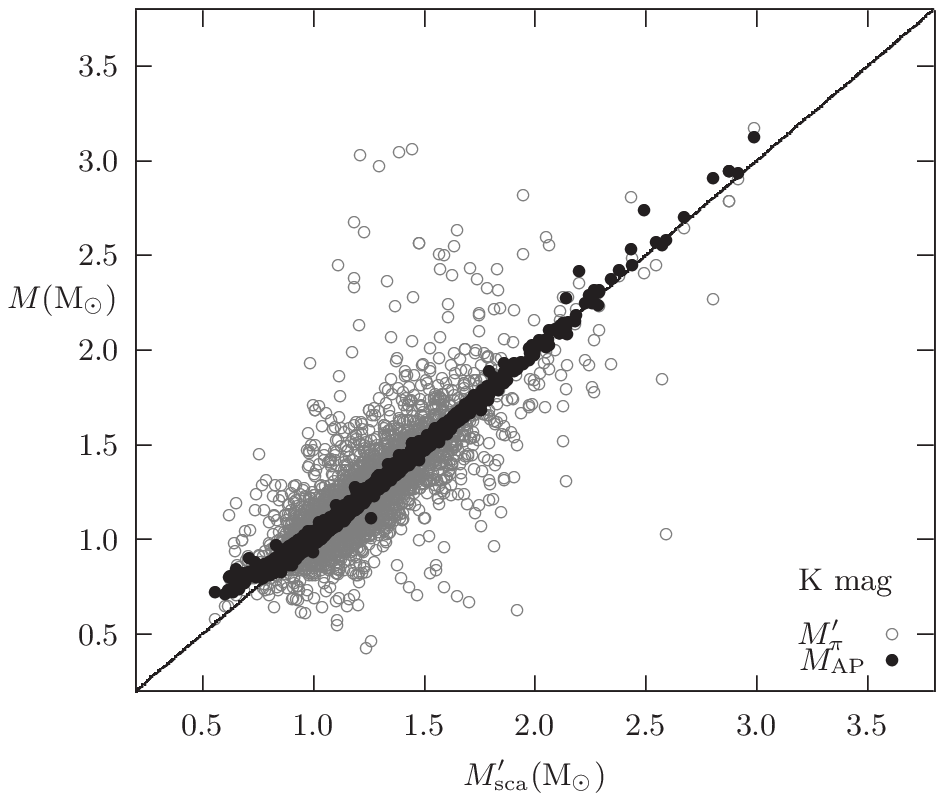}
\caption{The figure plots $M'_{\pi}$ and $M_{\rm AP}$ \wrt $M'_{\rm sca}$ for the $J$-, $H$- and $K$-band magnitudes.
The solid line denotes $M'_{\pi}=M'_{\rm sca}$. 
}
\end{figure}
{The mean difference between mass from the corrected parallax $M'_{\rm \pi}$ and mass $M_{\rm AP}$ given in APOKASC-2 is given in the last column of Table 1.
The difference is negligibly small for $K$ and $H$ magnitudes, about $-0.001$ and 0.010 \MS, respectively.
}

{Fig. 19 shows the first systematic comparison of {\it Gaia} and asteroseismic masses across a wide range in metallicity. 
Zinn et al. (2019b) compare the mass scales in the metal-poor regime, but not across all metallicities.
 In Fig. 19, $M'_{\rm \pi}$ and $M_{\rm AP}$ are plotted with respect to $M'_{\rm sca}$.
{$M_{\rm AP}$ is calibrated} to the dynamical mass scale from the NGC
6791 and 6819 clusters, and theoretical \Dnu~ corrections are applied. 
While there is an agreement between $M'_{\rm sca}$ and $M'_{\rm \pi}$ for the $J$ magnitude,  
a substantial difference is observed between $M'_{\rm sca}$ and $M_{\rm AP}$. For $H$- and $K$-band magnitudes, $M'_{\rm sca}$ and $M_{\rm AP}$ are in perfect agreement to the greatest extend. They are slightly different if $M<0.75$ \MS. 
}

{For some of the stars, fundamental properties are very accurate.
For 580 RGs, the difference between $M'_{\rm sca}$ and $M'_{\rm \pi}$ is less than 2 per cent.
The number of stars with mass difference less than 3 per cent is 837. A number of selected stars deserve to be studied in detail.
}

{Fig. 20 exhibits how the masses of RGs are distributed; 96 per cent of the stars have the mass range  0.8-1.8 \MS.  The radii of these stars range from 3.8 to 38 \RSbit.
The number of stars is the maximum at about 1.2 \MS.
}
\begin{figure}
\includegraphics[width=101mm,angle=0]{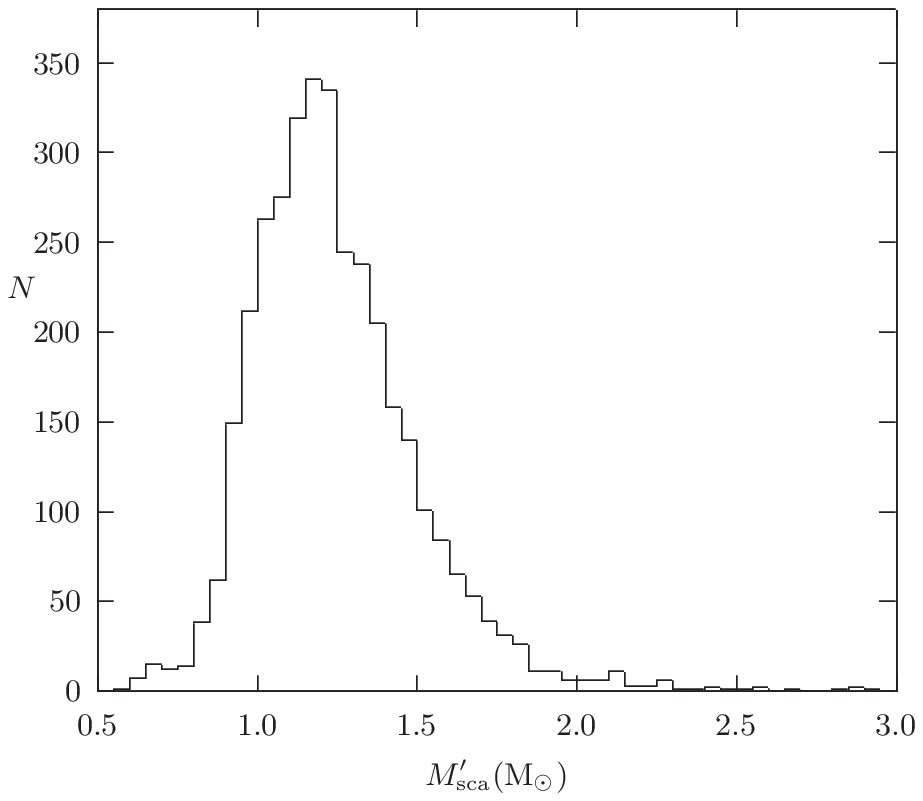}
\caption{Mass ($M'_{\rm sca}$) distribution of RGs.
}
\end{figure}

\section{Conclusions}
Combining all photometric, spectral, astrometric and asteroseismic data, we find the distance, radius and even mass of about 3600 stars from two different methods and compare these findings in order to assess {\it Gaia} parallaxes and to improve the asteroseismic scaling relations for stellar mass and radius. In this regard, the {\it Gaia}, {\it Kepler} and {\it TESS} {space missions} open a new era in our understanding of stellar evolution theory.

We first compare the first two data releases of {\it Gaia}. The DR2 parallaxes are more {precise} than the DR1 parallaxes. The 5 per cent {mean} difference between $d_{\rm sca}$ and $d_{\rm DR1}$
 (\yildiz et al. 2017) reduces to {2.8 per cent for $d_{\pi}$ for the Y16 DR2 (MS + SG + 2RGs) stars with $K$-band magnitude and without any correction. From the comparison of $R_{\rm sca}$ and $R_{\pi}$, we find {that $R_{\pi}=1.076 R_{\rm sca}^{0.942}$}. As there are only two RGs in this sample, we extend our study to the RGs in the {APOKASC-2} catalogue.
There are three samples of RGs we analyse. The luminosity of the stars is computed from the $V$ magnitude  for one group {(VRG) and from the {\it Gaia} $G$ magnitude for the other group (GRG)}. About 493 RGs are common in both groups. 
We also consider $J$-, $H$- and $K$-band magnitudes of RGs (JHKRG) in the APOKASC-2 catalogue. 
Almost for all of the RGs in the catalogue, $\sim$3500, their $J$-, $H$- and $K$-band magnitudes are available. 
From the comparison of $R_{\rm sca}$ and $R_{\pi}$, we obtain relationships for the VRG, GRG and JHKRG samples that are very similar to the relationship for the {Y16 DR2 sample}. }
These relationships  leads us to show how to improve the conventional scaling relations. Similar relationships are valid also for OEBs (solar-like oscillating RGs in eclipsing binaries).
We also find the relationship between $M_{\rm sca}$ and $M_{\pi}$. From the ratio  $\sqrt{(R_{\pi}/R_{\rm sca})^3/(M_{\pi}/M_{\rm sca})}$, we find the expression for $f_{\Delta \nu}$,
{assuming $f_{\nu_{\rm max}}=1$}. This expression for Y16 DR2 with $K$-band magnitude is almost identical to the one we have obtained from interior models for MS stars. 
{Interstellar extinction complicates the situation in most cases.} In this regard, we obtained the best solutions for the $H$- and $K$-band magnitudes.

Using these expressions for $f_{\Delta \nu}$, we can improve $M_{\rm sca}$ and $R_{\rm sca}$. Such a correction for $R_{\rm sca}$ and $M_{\rm sca}$ of the OEB stars means that there is agreement with radius and mass derived from orbital analysis. 

Improvement for $M$ and $R$ can also be done over $T_{\rm eff}$, because the dependences of asteroseismic ($d_{\rm sca}$, $R_{\rm sca}$ and $M_{\rm sca}$) 
and non- asteroseismic quantities ($d_{\pi}$, $R_{\pi}$ and $M_{\pi}$) on $T_{\rm eff}$ are completely different. While $R_{\rm sca}$, for example, 
is proportional to $T_{\rm eff}^{1/2}$, we find $R_{\pi}$ $\propto$ $T_{\rm eff}^{-2}$ for RGs with $T_{\rm eff}>4100$ K. Therefore, slight modifications in 
$T_{\rm eff}$ remove all the differences between asteroseismic and non-asteroseismic quantities for most of the stars. For the RGs with $T_{\rm eff}<4100$ K, 
we also take into account the derivative of $BC$ \wrt $T_{\rm eff}$. 
  
{
We also show how $f_{\Delta \nu}$, $f_{\nu_{\rm max}}$ and $\Delta \pi$ are inter-related (see equation 17). 
If we employ $f_{\Delta \nu}$ as derived for Y16 DR2 with $K$-band magnitude (equation 11) and take $f_{\nu_{\rm max}} =1$,
we obtain a parallax offset of about 0.005 mas.
}

{
We reach the best solution for RGs by using the model results of Sharma et al. (2016) for $f_{\Delta \nu}$.
Using the data of JHKRG, we show how $f_{\nu_{\rm max}}$ and $\Delta \pi$ are inter-related and we find that the difference between
$\pi - \Delta \pi$ and $\pi_{\rm sca}$ is $\pi$ dependent. {We fit a line to this trend for $\Delta \pi$ derived using $K$-band magnitudes} and we see that the 
slope vanishes at about $\Delta \pi=-0.0463$ mas for the $K$-band magnitude (see Table 1). At this value of $\Delta \pi$, the difference between $\pi - \Delta \pi$ and $\pi_{\rm sca}$ 
is very small. In this case, the mean value is $f_{\nu_{\rm max}}=1.003$. For the $H$ magnitude, 
we find that $f_{\nu_{\rm max}}=1.000$  and $\Delta \pi=-0.0424$. 
The use of these values of $f_{\nu_{\rm max}}$, $\Delta \pi$ and $f_{\Delta \nu\rm mod}$ from the models ensures 
that the asteroseismic and non-asteroseismic quantities are very compatible with each other (see Table 1), in particular for the $H$- and $K$-band magnitudes.
For these magnitudes, the asteroseismic mass $M'_{\rm sca}$ is in perfect agreement with the mass given in APOKASC-2 (see Fig. 19). 
The mean difference for the $K$-band magnitude, for example, is 0.001 \MS. 
We find that the mass and radius of the stars range from 0.8 to 1.8  \MS and from 3.8 to 38 \RS, respectively.
The number of stars is maximum at about 1.2 \MS.
}


	\section*{Acknowledgements}
We acknowledge discussions with Daniel Huber, M. Salaris and Antonio Frasca.
We would like to thank Dr. Frederick A. Magill for his professional checking of the language in the manuscript.
This work is supported by the Scientific and Technological Research Council of Turkey (T\"UB\.ITAK: 118F352).
~\\
~\\
        \section*{Data availability}
~\\
{The data underlying this article will be shared on reasonable request to the corresponding author.
} \\

\newpage

\appendix

\label{lastpage}

\end{document}